\documentclass[%
 reprint,
 amsmath,amssymb,
 aps,
prab,
]{revtex4-1}

\usepackage{graphicx}
\usepackage{subfigure}
\usepackage{dcolumn}
\usepackage{bm}
\usepackage{dcolumn}
\usepackage{subfig}
\usepackage{booktabs}
\usepackage{float}

\begin{document}

\preprint{APS/123-QED}

\title{Generation of large-bandwidth x-ray free electron laser with Evolutionary Many-Objective Optimization Algorithm}

\author{Jiawei Yan$^{1,2}$} 
\author{Haixiao Deng$^{1,}$}%
\email{denghaixiao@sinap.ac.cn}  \affiliation{%
	$^1$Shanghai Institute of Applied Physics, Chinese Academy of Sciences, Shanghai 201800, China\\ $^2$University of Chinese Academy of Sciences, Beijing 100049, China. 
}%

\date{\today}
\begin{abstract}
X-ray free-electron lasers (XFELs) are cutting-edge scientific instruments for a wide range of disciplines. Conventionally, the narrow bandwidth is pursued in an XFEL. However, in recent years, the large-bandwidth XFEL operation schemes are proposed for X-ray spectroscopy and X-ray crystallography, in which over-compression is a promising scheme to produce broad-bandwidth XFEL pulses through increasing the electron beam energy chirp. In this paper, combining with the beam yaw correction to overcome the transverse slice misalignment caused by the coherent synchrotron radiation, finding out the over-compression working point of the linac is treated as a many-objective (having four or more objectives) optimization problem, thus the non-dominated sorting genetic algorithm III is applied to the beam dynamic optimization for the first time. Start-to-end simulations demonstrate a full bandwidth of 4.6\% for Shanghai soft x-ray free-electron laser user facility.

\end{abstract}

\maketitle

\section{Introduction}
X-ray free-electron lasers (XFELs) can provide short wavelength radiation with high brightness and ultra-fast time structures. They are leading-edge instruments in a wide range of research fields \cite{1}. Most worldwide XFEL facilities are based on self-amplified spontaneous emission (SASE) \cite{2,4,5} process and the relative bandwidth at saturation is between ${10^{ - 3}}$ and ${10^{ - 4}}$ \cite{6}. To generate fully coherent XFEL pulses, self-seeding \cite{48,49}, external seeding \cite{7,8,9} and other techniques \cite{50,51,52} are used for further decreasing the XFEL bandwidth. However, besides narrow bandwidth FEL pulses, the large-bandwidth FEL operation has attracted increasing attentions \cite{10}. Broad-bandwidth FEL pulses are very useful in many spectroscopy experiments \cite{11,12,13,14} and X-ray crystallography \cite{11,15,16}. Furthermore, the large-bandwidth operation mode allows for FEL wavelength tuning in a more flexible way. 

According to the FEL resonance condition \cite{17}, the wavelength of FEL radiation is determined by the electron beam energy and the undulator field parameters. In principle, properly sending the head-tail tilted electron bunches into a transverse gradient undulator \cite{22} or into a planar undulator with natural gradient \cite{23} can make different parts of the bunch experience different magnetic field, and thus generate broadband FEL pulses. Besides that, using electron beams with time-energy correlation is a more natural way to obtain broad-bandwidth FEL, which may be achieved without additional hardware elements in currently existed facilities. The simplest way to obtain energy chirp is off-crest acceleration, while it is inefficient and at the cost of reducing beam energy. There is another special compression mode named over-compression \cite{19,20,21} which can be used to generate a large energy chirp. In this scheme electron beams are over-compressed in the bunch compressor which means the head and tail of the bunch will interchange their positions. The sign of energy chirp is changed too. Thus, the beam energy chirp will be further increased by the wakefields of subsequent rf structures. The crucial issue of over-compression mode is to find appropriate working point of the linac in which the electron beams have large energy chirp while other beam qualities can be maintained. The above process can be treated as a many-objective (having four or more objectives) optimization problem in which the energy chirp, peak current, energy spread, and current profile of the electron bunch are objectives. 

Pareto-dominance based multi-objective evolutionary algorithms (MOEAs), such as non-dominated sorting genetic algorithm II (NSGA-II) \cite{27}, have been widely and successfully used in the accelerator community \cite{28,29,30,31,32} for physics study and optimization. However, in recent years, it has been pointed out that pareto-dominance based MOEAs will encounter some difficulties in many-objective optimization problems (MaOPs) \cite{33}. The main reason is the severe loss of selection pressure towards the Pareto front which is caused by an increase in the objectives. Moreover, the diversity-preservation operator like crowding distance operator will become time consuming in many-objective problems. It is a common strategy to transform multiple objectives into one or two objectives by using the weighted method. Nevertheless, not only it is difficult to determine weights in the scenario, but it also loses the opportunity to analyze the relationships between each objective.

Recently, an improved NSGA-II procedure, which was termed NSGA-III \cite{34,42}, has been proposed as an evolutionary many-objective algorithm. NSGA-III maintains the diversity among population members by supplying and adaptively updating a set of well-spread reference points. It has been demonstrated that NSGA-III is efficient in optimizing 2 to 15 objectives problems \cite{44,45,46}. In this paper, using the Shanghai soft x-ray free-electron laser (SXFEL) user facility parameters, the over-compression process is optimized by adjusting accelerator operation parameters with NSGA-III to explore the maximum FEL bandwidth. In addition, over-compressed electron bunches have gone through full compression status in chicane where the coherent synchrotron radiation (CSR) becomes quite strong \cite{47}. The slice misalignment along the electron bunch caused by the CSR leads to a projected emittance growth, which will be hindered for the broadband FEL lasing. More recently, dispersion section based beam yaw correction has been proposed and experimentally verified \cite{25,35}. However, the current profile and energy chirp of electron beam may be changed after such a correction. Therefore, in this study, the beam yaw correction is also included into the evolutionary many-objective optimization.  

The paper is structured as follows. In Sec.\ II, the optimization strategy of the over-compression mode for SXFEL user facility is described, including the objectives and the algorithm used. The optimization results of NSGA-III and three typical cases are shown in Sec.\ III. The many-objective optimization including the beam yaw correction is presented in Sec.\ IV. The conclusions are summarized in Sec.\ V.

\section{Optimization strategy}

To obtain high quality electron beams in the over-compression mode, lattice parameters including the voltages and phases of the accelerating sections and the angels of the bunch compressors in the linac are optimized. The corresponding optimization objectives consist of the peak current, slice energy spread, current profile, and energy chirp of the electron bunch at the entrance of the undulator. In this optimization, the start-to-end simulations are performed to compute these optimization objectives and validate the large-bandwidth XFEL generation in the undulator. ASTRA \cite{53} is used to track electron beams in the injector where the transverse space charge forces are strong. Tracking simulation in the main linac is performed by the ELEGANT \cite{54} in which collective effects like the CSR, longitudinal space charge and wakefields are considered. The objectives calculation are based on the ELEGANT simulation results. Thus, to balance the calculation accuracy and time spent, the ELEGANT simulation is performed with one hundred thousand macroparticles during the optimization. GENESIS \cite{55} is used to verify the XFEL generation. 

An electron bunch from the ELEGANT simulation result is divided into 100 slices to calculate objective values. The peak current $I_{max}$ is defined as the maximum current value of these slices. 2\% of the $I_{max}$ is chosen as a cut-off point. The cut-off points on two sides of a bunch are a and b respectively. The objective value of the energy spread, $\delta _{mean}$, is defined as average slice energy spread of slices between the two cut-off points:
\begin{equation}
{\delta _{mean}}{\rm{  =  }}\frac{{\sum\limits_a^b {{\delta _i}} }}{{b - a}},
\end{equation}
where $\delta _i$ represents the energy spread of the i-th slice. The energy chirp is defined as the relative energy difference between the two cut-off points:
\begin{equation}
{\sigma _d}{\rm{  = }}\frac{{{\rm{ }}\left| {\left. {{\gamma _b} - {\gamma _a}} \right|} \right.}}{{\frac{1}{{b - a}}\sum\limits_a^{\rm{b}} {{\gamma _i}} }},
\end{equation}
where $\gamma_i$ is the energy of the i-th slice. To discribe the current profile of an electron beam, a profile factor, $C$, is defined as the ratio of the sum of central 50 slice current to the sum of total slice current:
\begin{equation}
C = \frac{{\sum\limits_{26}^{75} {{I_i}} }}{{\sum\limits_1^{100} {{I_i}} }}.
\end{equation}
It should be pointed out that the profile factor does not describe a specific shape. This is to analyze the relationship between current shape and energy chirp during the optimization.

 These optimization goals are not independent of each other. For example, peak current and slice energy spread are two conflicting objectives. Besides that, the current profile will influence the energy chirp brought by longitudinal wakefields. Therefore, it is important for the over-compression mode to optimize all the objective at the same time. As aforementioned, those widely used MOEAs are originally proposed for problems with two or three objectives. In recent years, numerous evolutionary many-objective optimization algorithms have been put forward to handle those problems with more than three objectives. The NSGA-III, one of the most frequently-used many-objective optimization algorithm, is applied to optimize the over-compression mode in SXFEL user facility.

\begin{figure*}[htp] 
	\centering 
	\includegraphics[width=\linewidth]{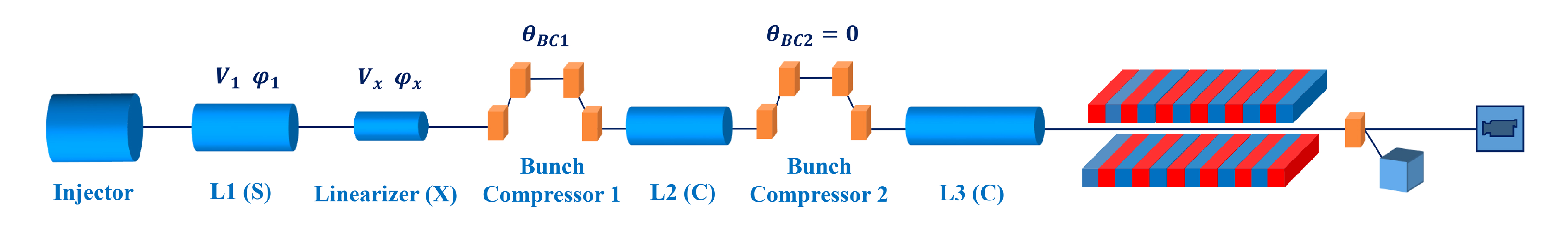}
	\captionsetup{justification = raggedright, singlelinecheck = false}
	\caption{Layout of the SXFEL user facility linac, with the injector section, S-band sections (L1), one X-band linearizer (LX), C-band sections (L2, L3) and two bunch compressor chicanes (BC1, BC2). Angle of the BC1, voltages and phases of L1 and LX are optimized for the large-bandwidth operation mode. The second bunch compressor is turned off to utilize more longitudinal wakefields.}
	\label{FIG1} 
\end{figure*}

The NSGA-III is an improved version of the NSGA-II for MaOPs. The basic framework of NSGA-III is similar with the NSGA-II. Take the t-th generation as an example. Assume the population is $Q_{t}$ and its size is N. After the tournament selection, polynomial mutation, and simulated binary crossover, an offspring $P_{t}$ is generated from $Q_{t}$ with the same size. Thereafter, the $Q_{t}$ and $P_{t}$ are merged to $R_{t}$, and the best N members need to be selected from the $R_{t}$ for the next generation. The selection in NSGA-III consists of two stages, one stage is used to guarantee the convergence to Pareto front and the other stage maintains the diversity of population. The first stage is the same as the NSGA-II which is based on the Pareto dominance to sort the population into multiple non-dominated fronts. If the total number of individuals in the top M Pareto fronts is larger than the population size, the first M-1 Pareto fronts will be selected and the rest individuals are chosen from the M-th Pareto front based on the diversity keeping methods. In NSGA-II, solutions in the the M-th Pareto front with largest crowding distance will be chosen. However, the crowding distance measure is not working well for the MaOPs. In NSGA-III, the diversity is preserved by utilizing a set of well-distributed reference points. In this selection mechanism, objectives and the reference points are normalized to be in the same range. Thereafter, every population member is associated with a reference point based on a perpendicular distance measure. Next, to ensure the diversity, some population members are accepted after a niche operation. The reference points in the optimization of the over-compression mode are based on the systematic approach proposed by Das and Dennis \cite{41} treating all objectives equally.

\section{Over-compression mode in SXFEL user facility}

As the first X-ray FEL in China, the SXFEL user facility is under construction at Shanghai \cite{43}. The SXFEL user facility will be equipped with two undulator lines. One is the two-stage seeded FEL line to generate 3 nm fully coherent FEL pulse, and the other is SASE line aimed at 2 nm on the basis of in-vacuum undulator. The scope of the optimization in this study is to design the large-bandwidth operation mode for the SASE line.

\begin{figure*}  
	\begin{minipage}[htp]{0.5\linewidth}  
		\centering  
		\includegraphics[width=0.9\columnwidth]{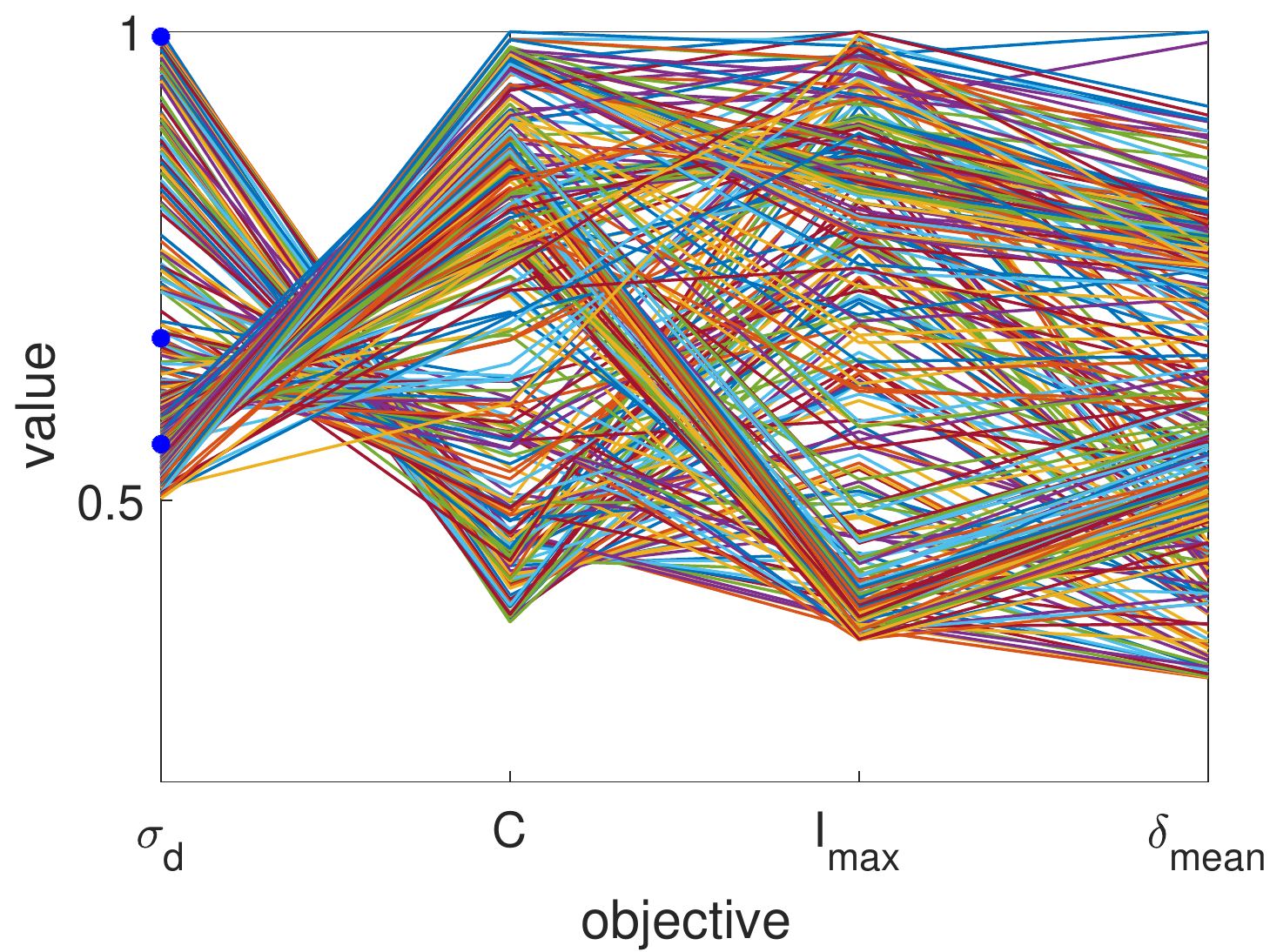}  
	\end{minipage}%
	\begin{minipage}[htp]{0.5\linewidth}  
		\centering  
		\includegraphics[width=0.9\columnwidth]{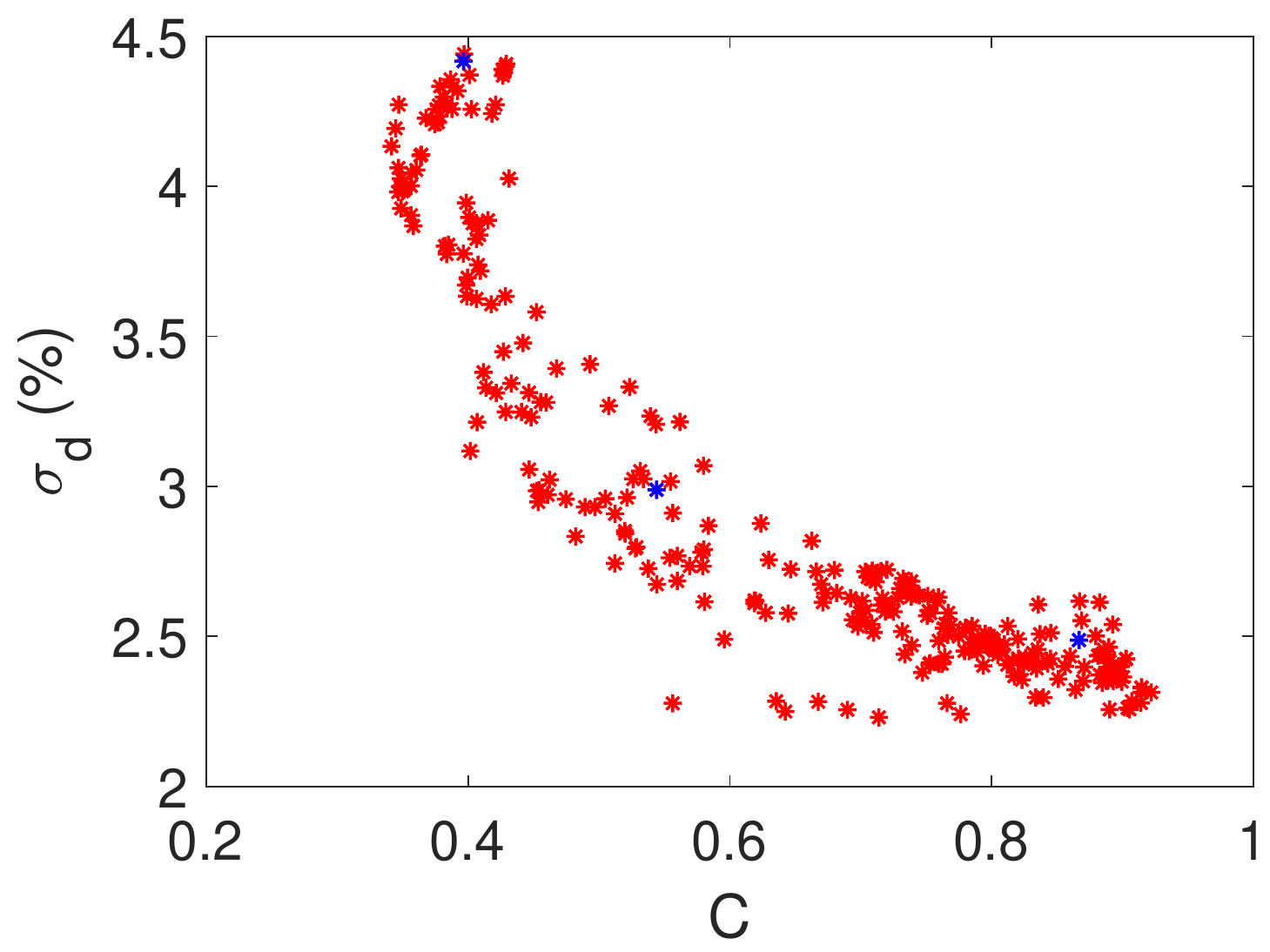}  
	\end{minipage} 
	\captionsetup{justification = raggedright, singlelinecheck = false}
	\caption{Parallel coordinate plots of the Pareto-optimal front for the optimization of the four objectives with the NSGA-III (left). Projection of the Pareto front on the plane formed by the energy chirp and current profile factor (right). The three blue dots show three typical solutions whose current profiles and longitudinal phase spaces are presented in Fig.\ \ref{FIG3}.}
	\label{FIG2} 
\end{figure*}

In the baseline design of the SXFEL user facility, 0.5 nC electron bunches with 130 MeV are generated in the injector section which includes two S-band accelerating structures and a laser heater. The downstream of the injector is the main linac with three accelerating sections and two bunch compressors. Electron beams are accelerated at the off-crest phase of a S-band accelerating section to create an energy chirp and an X-band rf cavity is used to linearize the chirp. Following this, electron beams with 256 MeV are compressed in the first magnetic chicane. There are two C-band linac accelerators which are used to further increase the energy to 1.5 GeV and the second chicane is between them. Finally, electron bunches are sent to the two FEL lines. In the over-compression mode, energy chirp is increased by the longitudinal wakefields after the over-compression. Therefore, it is more appropriate to make the electron beam be over-compressed in the first bunch compressor and turn off the second bunch compressor to utilize more longitudinal wakefields. Layout of the main linac with a single-stage bunch compressor and the SASE line are shown in Fig.\ \ref{FIG1}.

\begin{figure*}  
	\begin{minipage}[htbp]{0.33\linewidth}  
		\centering  
		\includegraphics[width=\columnwidth]{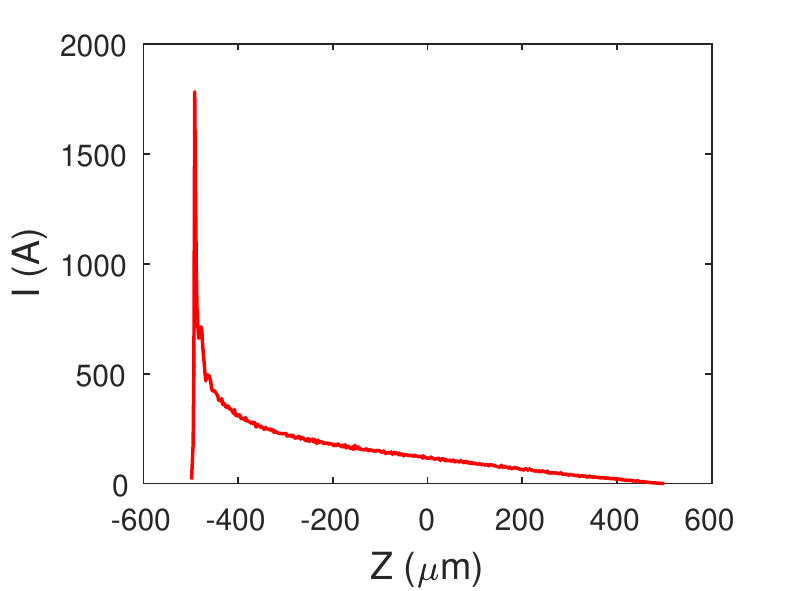}  
	\end{minipage}%
	\hfill 
	\begin{minipage}[htbp]{0.33\linewidth}  
		\centering  
		\includegraphics[width=\columnwidth]{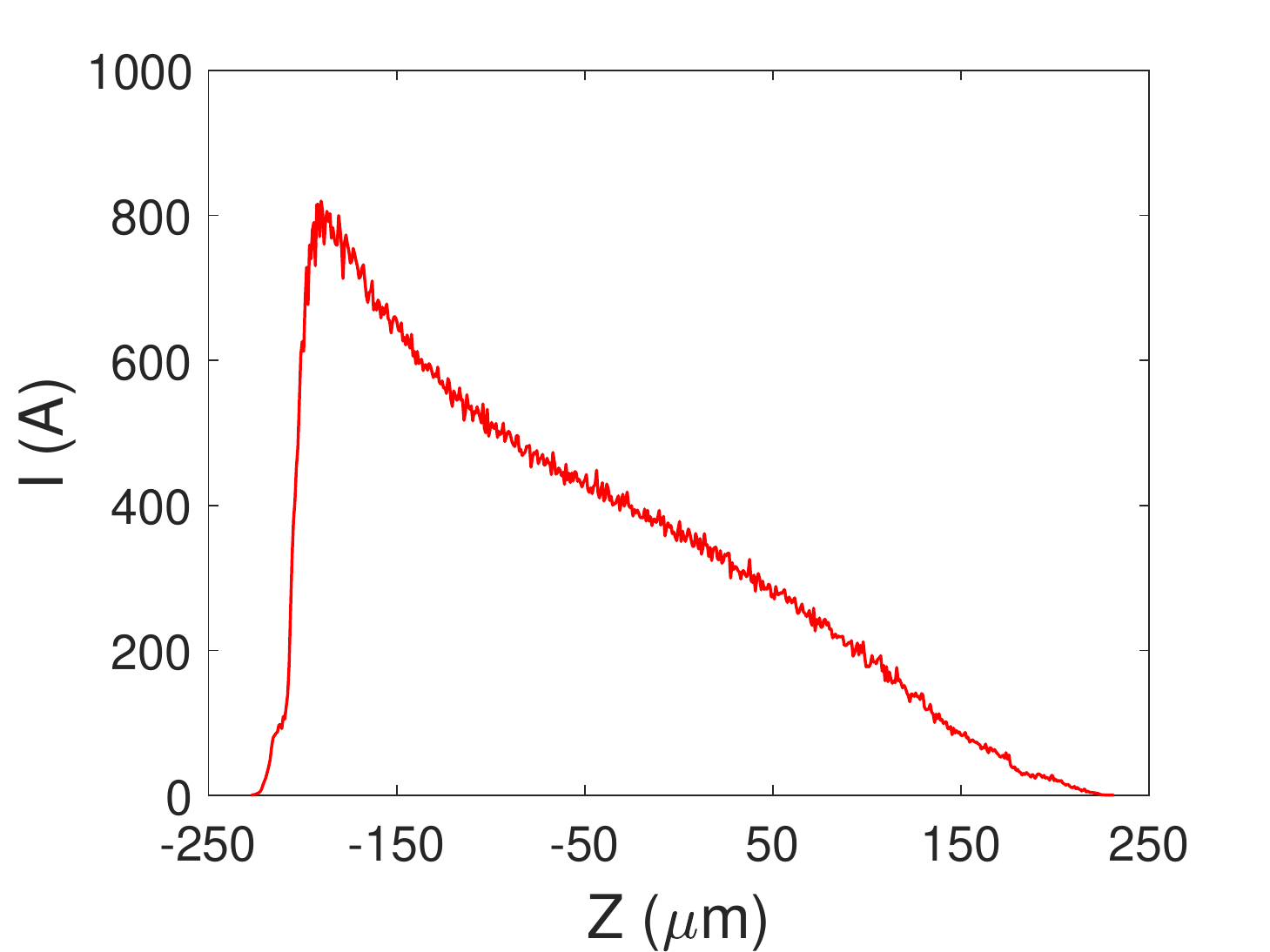}  
	\end{minipage}%
	\hfill 
	\begin{minipage}[htbp]{0.33\linewidth}  
		\centering  
		\includegraphics[width=\columnwidth]{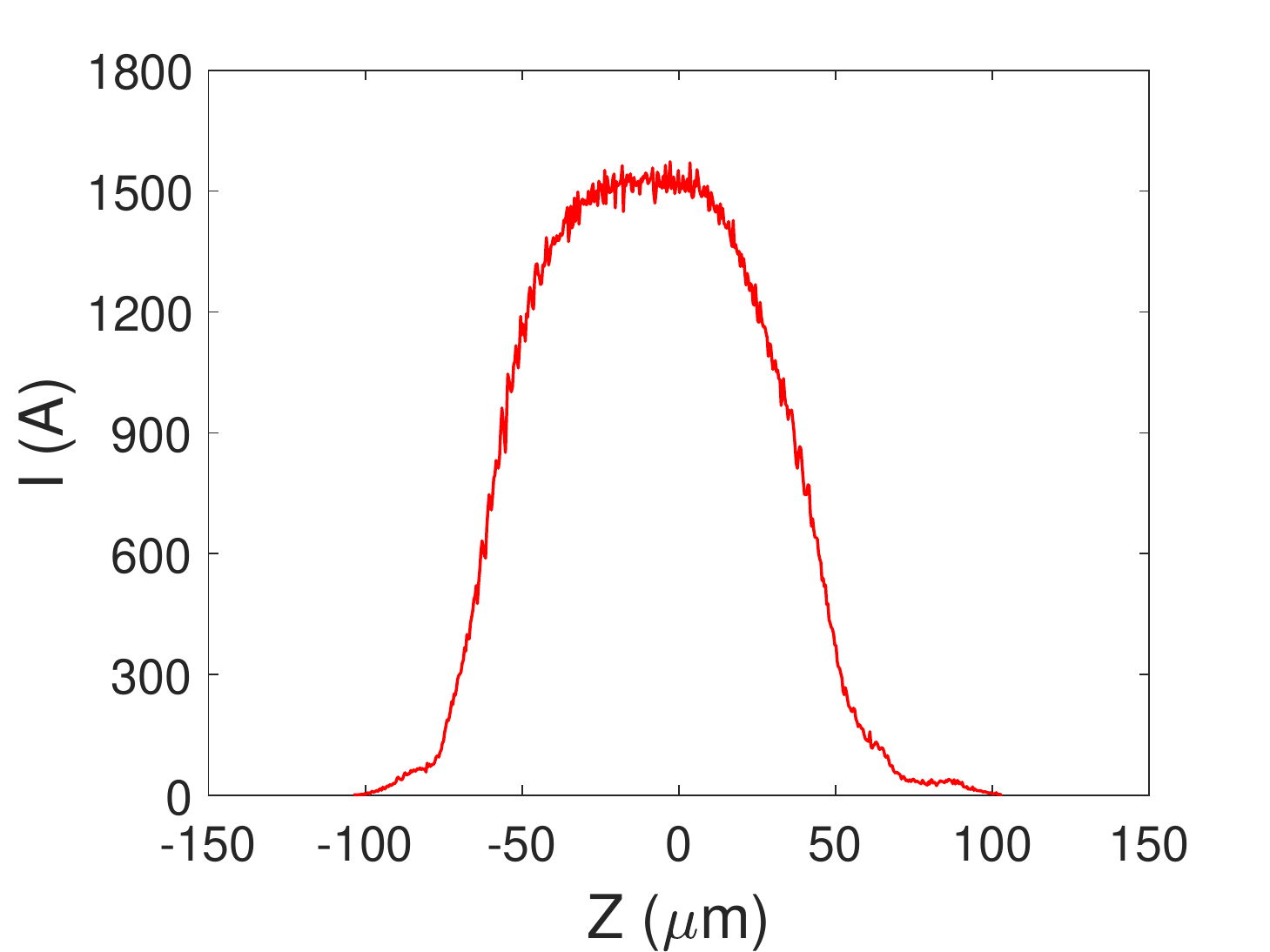}  
	\end{minipage}%
	
	\hfill 
	\begin{minipage}[htbp]{0.33\linewidth}  
		\centering  
		\includegraphics[width=\columnwidth]{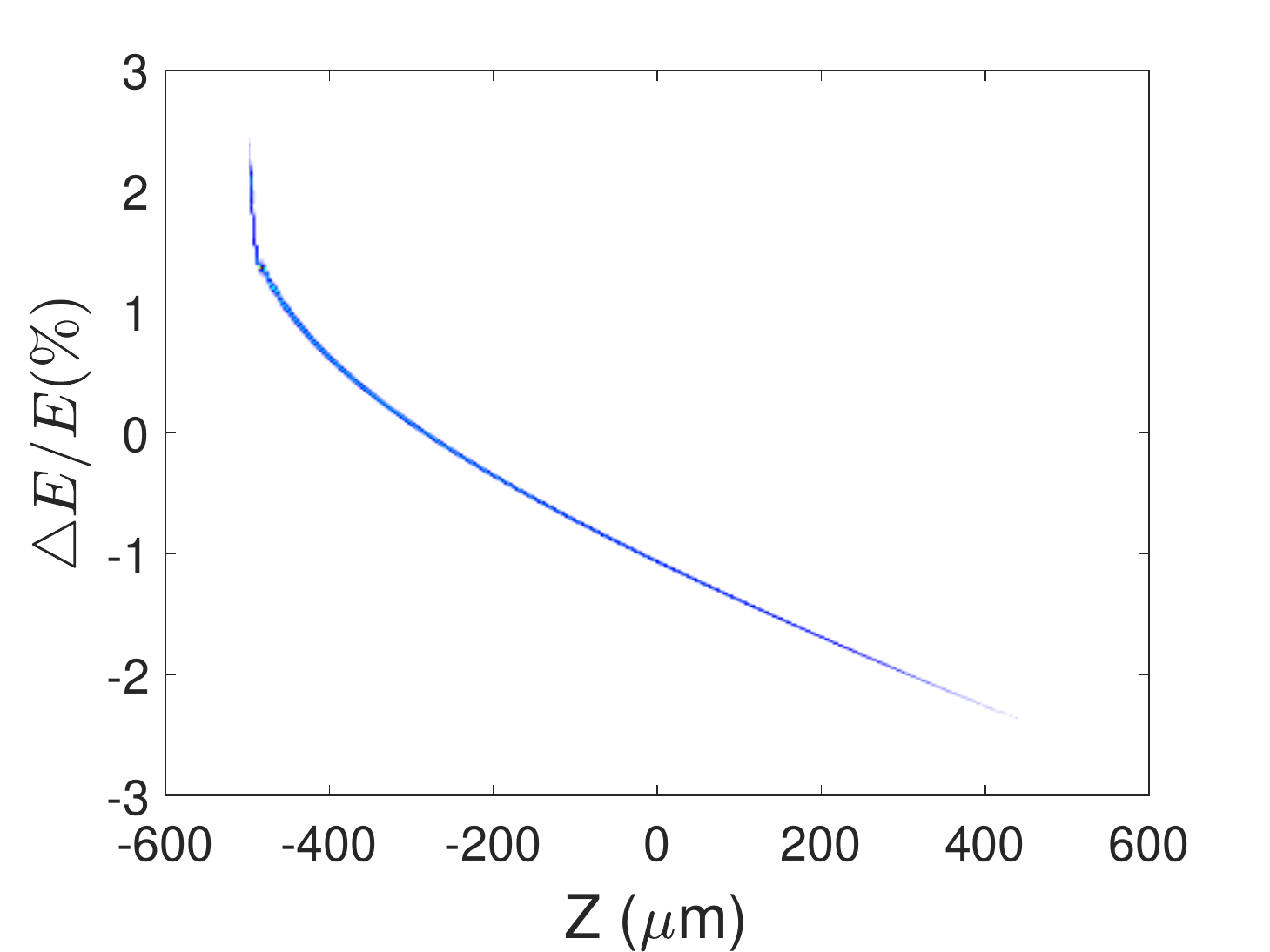}  
	\centerline{(a) case 1}
	\end{minipage}%
	\hfill 
	\begin{minipage}[htbp]{0.33\linewidth}  
		\centering  
		\includegraphics[width=\columnwidth]{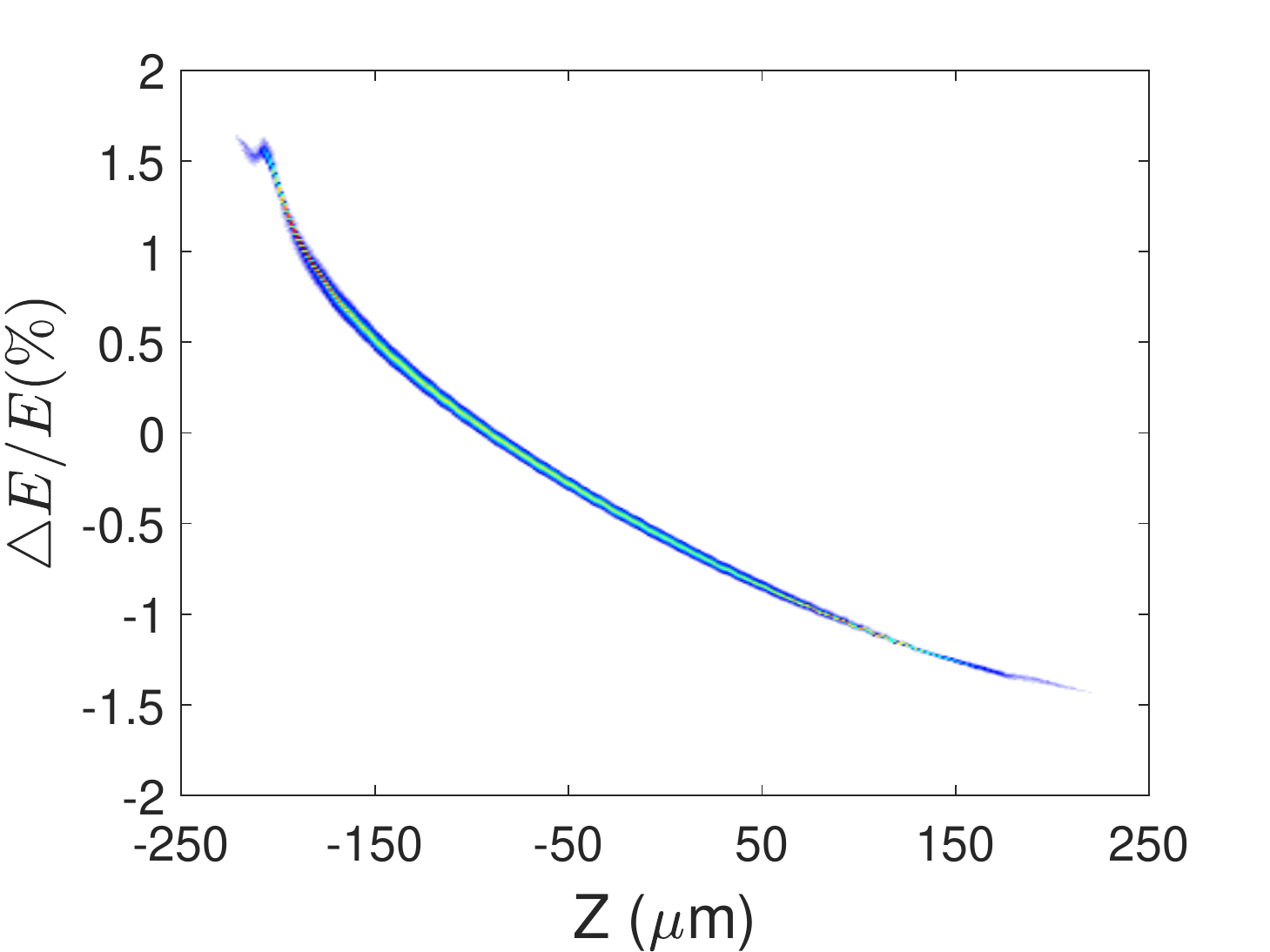}  
		\centerline{(b) case 2}  
	\end{minipage}%
	\hfill 
	\begin{minipage}[htbp]{0.33\linewidth}  
		\centering  
		\includegraphics[width=\columnwidth]{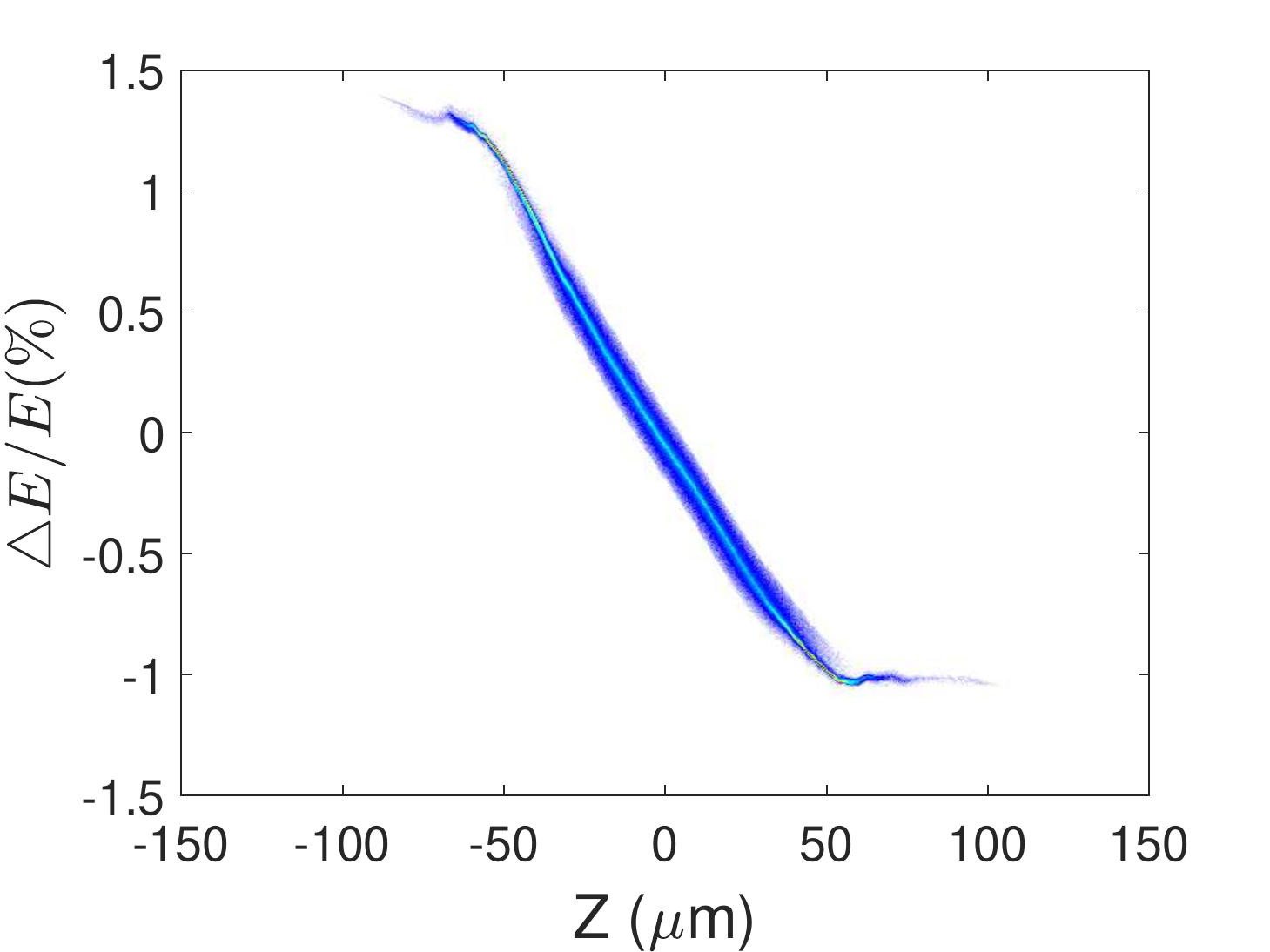}  
		\centerline{(c) case 3}
	\end{minipage}%
	\captionsetup{justification = raggedright, singlelinecheck = false}
	\caption{Current profile (upper) and longitudinal phase space (lower) of the three typical electron bunches selected from the Pareto-optimal front (see the three blue dots in Fig.\ \ref{FIG2}).}
	\label{FIG3} 
\end{figure*}

The angle of the first bunch compressor ($\theta_{BC1}$), voltages and phases of the S-band accelerating section ($V_1$, ${\varphi _1}$) and the X-band linearizer ($V_x$, ${\varphi _x}$) are chosen as optimization variables. The ranges of these variables are decided by the limits of related hardware. Besides the limitations of variables, there are some restrictions on the beam qualities for FEL lasing, for instance, the slice energy spread. If those solutions with poor beam qualities are discarded earlierly, it will help to improve the efficiency of the optimization. Thus, those solutions with a peak current more than 2000 A or less than 700 A will be given the worst fitness in this optimization. In addition, those electron bunches which are not over-compressed, i.e., the tail of the electron bunch with larger energy than the head, will also be given the worst fitness value for faster convergence. The population and iteration in NSGA-III are set as 300 and 100. The Pareto front of the final iteration are shown in Fig.\ \ref{FIG2} (left) where objectives are normalized. For normalization, values of energy chirp, profile factor, peak current and slice energy spread are devided by 4.44\%, 0.92, 2000 A, and $5.62\times10^{-4}$ separately. Fig.\ \ref{FIG2} (right) shows the projection of the Pareto front on the plane formed by the profile factor and the energy chirp. In general, a worse profile factor can achieve a larger energy chirp. From the prospective of the current shape, solutions can be divided into three categories. And three typical cases from the three kinds of current profile are given in Fig.\ \ref{FIG2} (right) with blue dots. The current profile and longitudinal phase space of the three cases are presented in Fig.\ \ref{FIG3}. 

The first kind of electron beams are those with profile factor lower than 0.5, which have a high-current leading peak and a long tail like the case 1 in Fig.\ \ref{FIG3} (a). The sharp single-horn of this kind of electron beam is caused by the nonlinear bunch compression. The maximum energy chirp available in this type of electron beams is 4.4\%. Although the energy chirp is large, the low-current tail of such kind of electron beams will not contribute to the broad-bandwidth FEL lasing. Therefore, this kind of electron beams are not suitable for the large-bandwidth operation mode. Furthermore, such sharp current profile has been proposed to produce single-spike x-ray pulses \cite{36,37}. It can further shorten the pulse by combining the energy chirp with an optimized reverse taper undulator. When the purpose is to generate the short pulses, it just need change corresponding optimization objectives of the NSGA-III like the length and energy chirp of the high-current part. 

The second kind is those electron beams with profile factor between 0.5 and 0.7. This kind of electron bunches have a quasi-triangular current profile like the case 2 in Fig.\ \ref{FIG3} (b) with a profile factor of 0.54. The energy chirp of the case 2 is 3\% which is the largest value in this kind. Similar to the first kind of electron beams, the low-current tail does not contribute to the FEL lasing which causes the final FEL bandwidth to be much smaller than the theoretical value, i.e. the twice of the energy chirp. In addition, this kind of current shape also leads to a poor uniformity of the power profile. Thus, this kind of electron bunches are not the best choice for producing broad-bandwidth XFEL pulses. Nevertheless, this kind of electron bunches with triangular shape are capable of improving the transformer ratio in the beam-driven collinear wakefield accelerators \cite{38,39}. For such applications, the current values of each slice can be treated as optimization targets, to obtain a more precise triangular shape.

\begin{figure}  
	\begin{minipage}[htp]{0.5\columnwidth}  
		\centering  
		\includegraphics[width=\columnwidth]{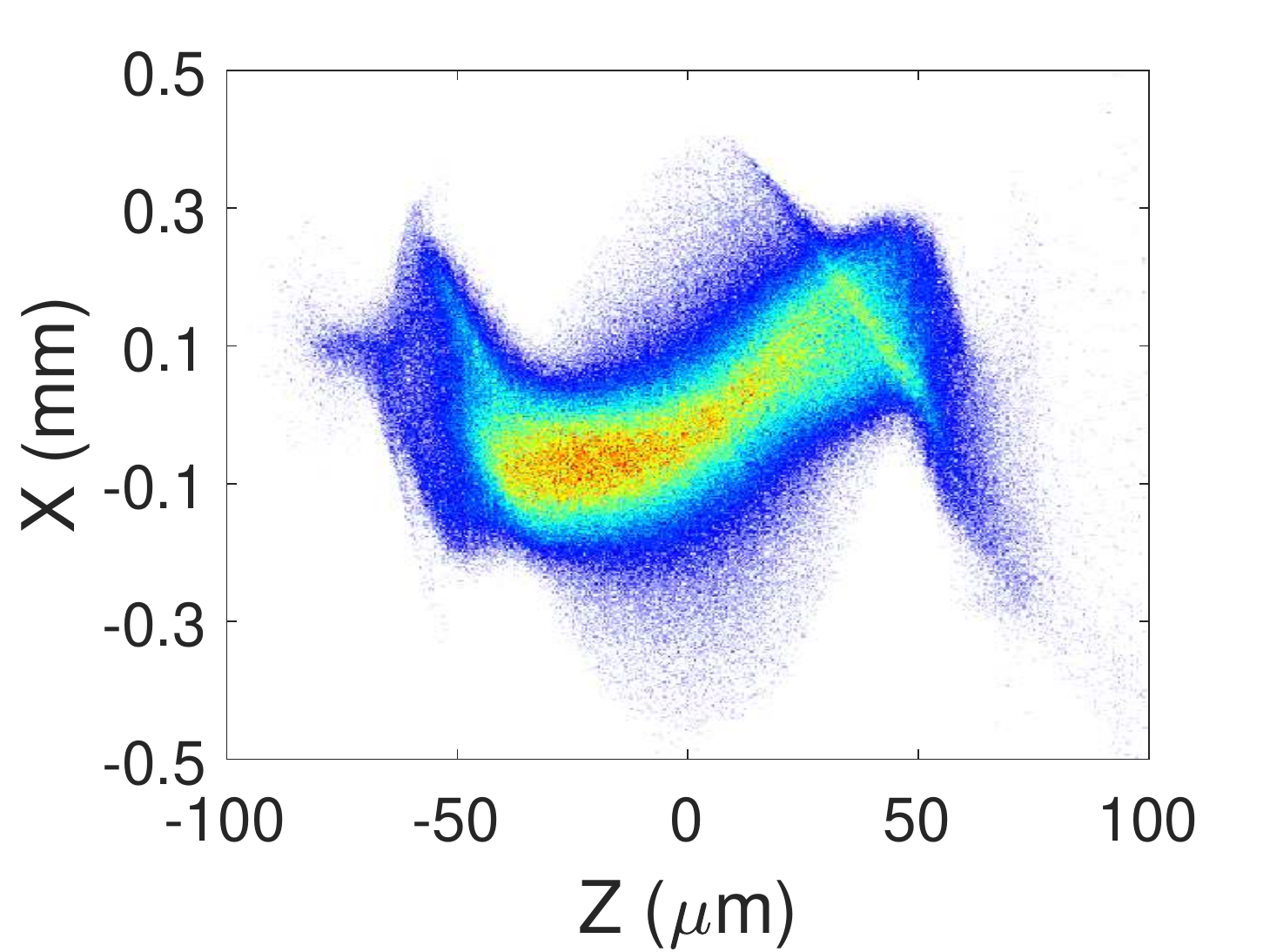}  
	\end{minipage}%
	\begin{minipage}[htp]{0.5\columnwidth}  
		\centering  
		\includegraphics[width=\columnwidth]{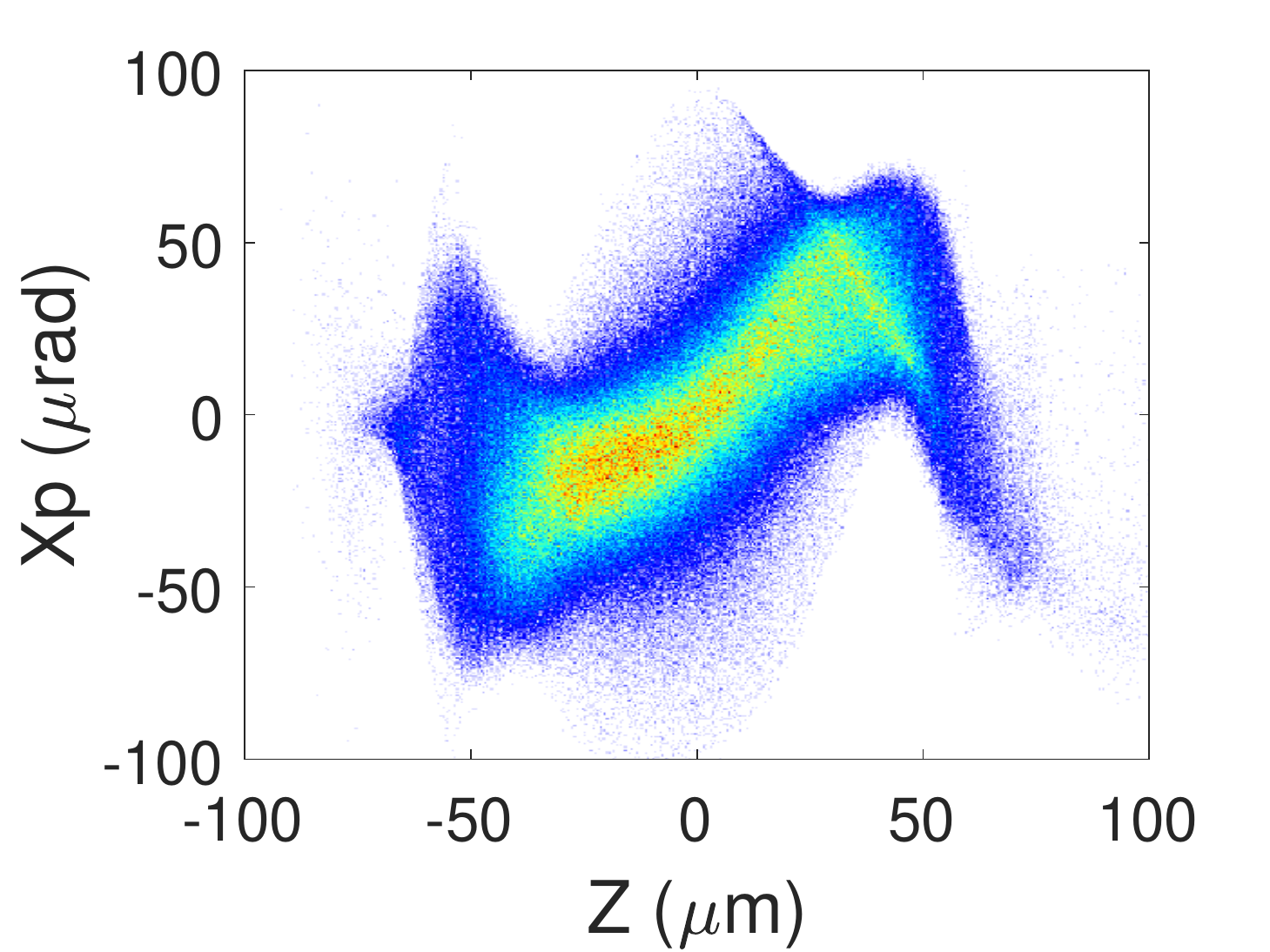}  
	\end{minipage}  

		\begin{minipage}[htp]{0.5\columnwidth}  
		\centering  
		\includegraphics[width=\columnwidth]{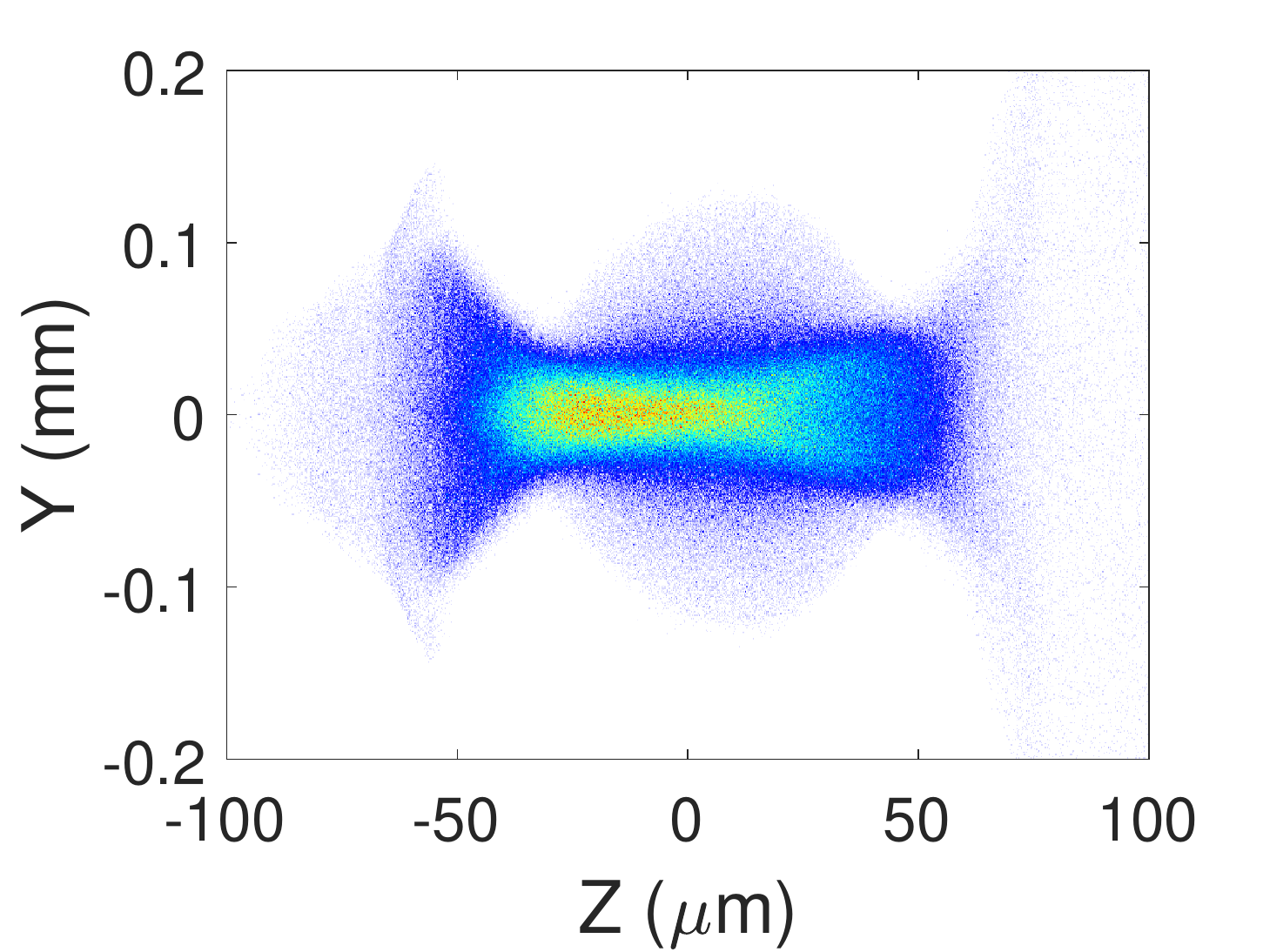}  
	\end{minipage}%
	\begin{minipage}[htp]{0.5\columnwidth}  
		\centering  
		\includegraphics[width=\columnwidth]{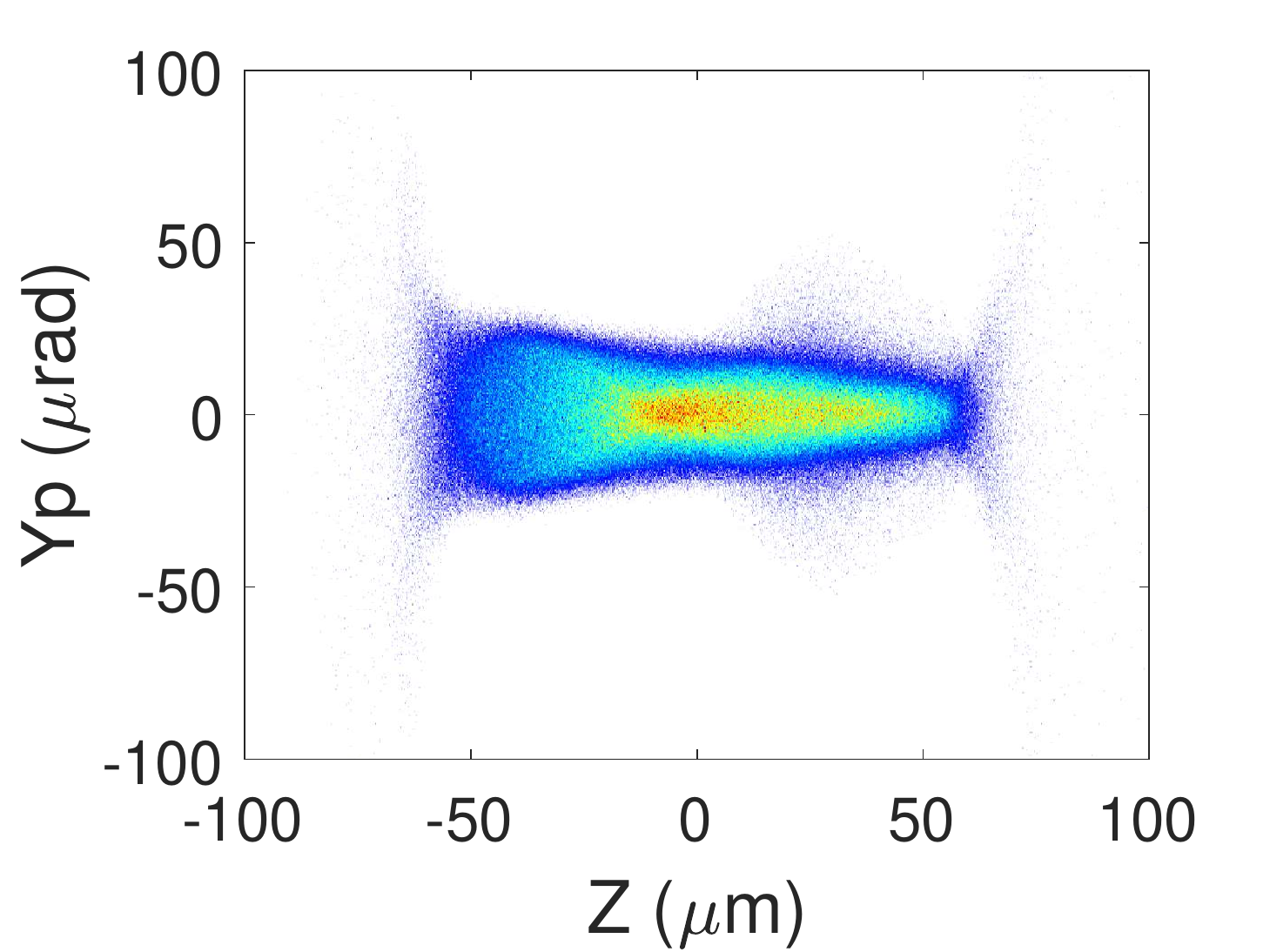}  
	\end{minipage}  
    \captionsetup{justification = raggedright, singlelinecheck = false}
	\caption{Transverse phase space of the case 3 at the entrance of the undulator.}
	\label{FIG4} 
	
\end{figure}  

Those electron beams with a profile factor more than 0.7 are treated as the third kind. The current shape of this kind of electron bunches is more or less flat-top or Gaussian shape, which are suitable for generating large-bandwidth FEL pulses to most users. The energy chirp and profile factor of the case 3 (see Fig.\ \ref{FIG3} (c)) are 2.39\% and 0.89. Start-to-end simulations with one million macro particles have been performed based on the rf parameters of this case. Ideally, it will produce radiation with a bandwidth that is close to 4.8\%. However, in this case, the full width FEL bandwidth including a 2\% cut is 3.1\%. The main reason for this bandwidth is the strong CSR in the over-compression mode. When the CSR effects are not considered in the ELEGANT simulation, the final FEL bandwidth is 4.5\%. The CSR brings electron bunches with a longitudinally dependent energy loss which will be turned into slice misalignment called beam yaw and increase the projected emittance of the bending plane. The beam yaw is further increased in the subsequent accelerating sections due to the transverse wakefields. Fig.\ \ref{FIG4} shows the transverse phase space of the case 3 at the entrance of the undulator, in which there is a beam yaw in the horizontal phase space. Therefore, eliminating the CSR caused beam yaw is important for the large-bandwidth FEL generation.

\begin{figure}
	\centering  
	\includegraphics[width=\columnwidth]{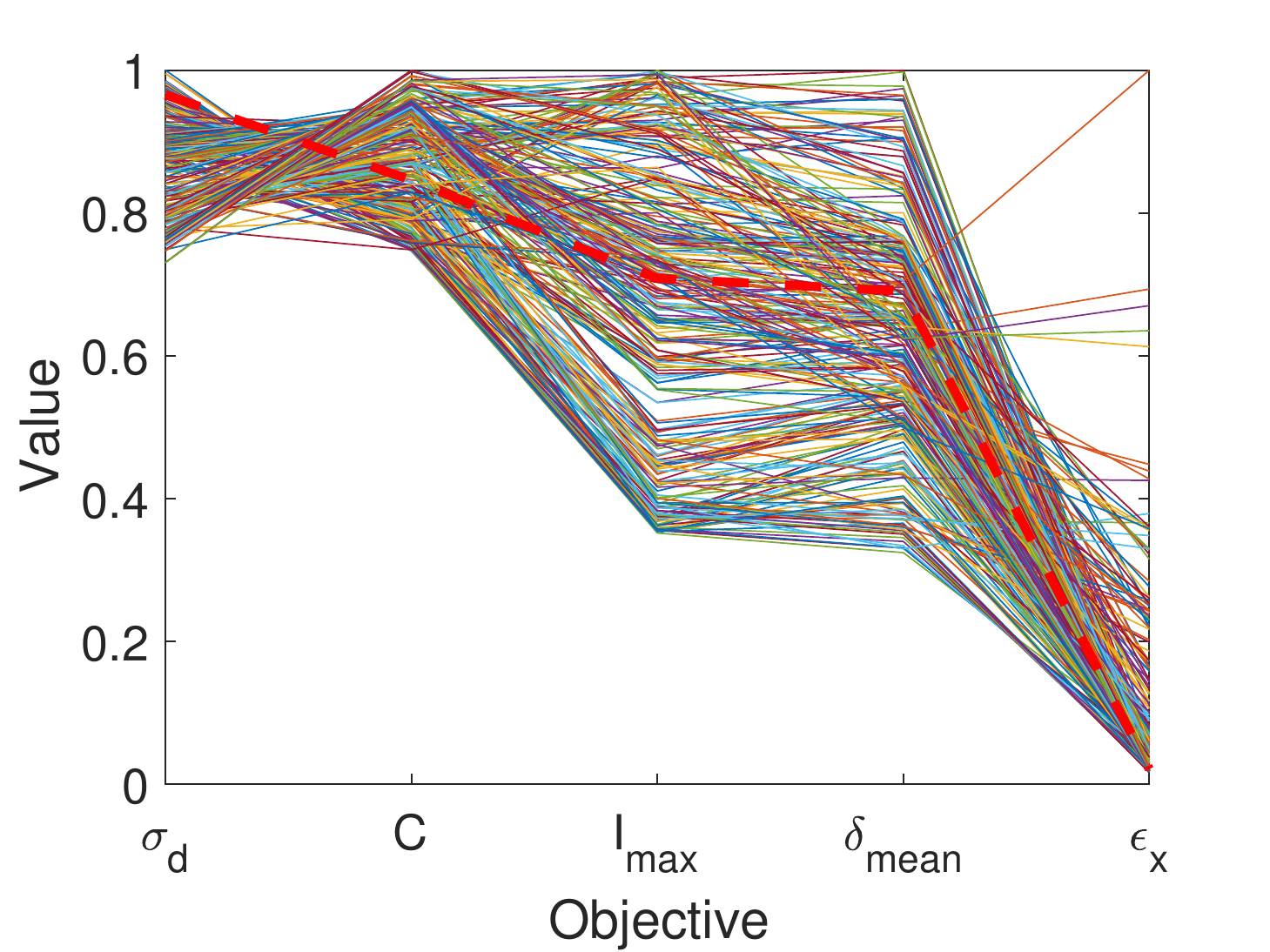}  
	\captionsetup{justification = raggedright, singlelinecheck = false}
	\caption{The Pareto-optimal front for the optimization of the five objectives. The red dashed line indicates the chosen case whose current profile and longitudinal phase space are shown in Fig.\ \ref{FIG6}.}
	\label{FIG5} 
\end{figure}

\section{Combining with beam yaw correction}

\begin{figure}  
	\begin{minipage}[htp]{0.5\linewidth}  
		\centering  
		\includegraphics[width=\columnwidth]{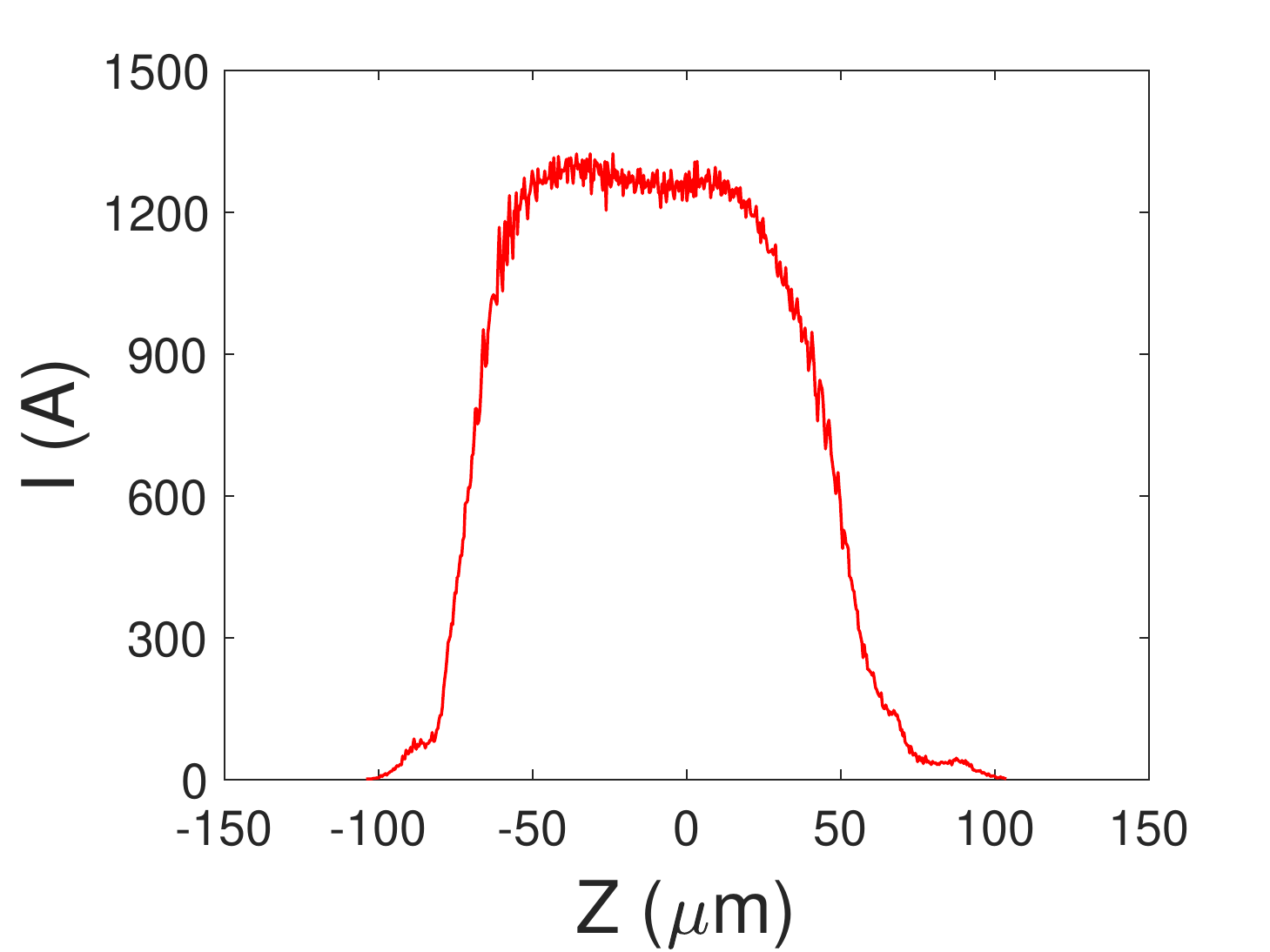}  
	\end{minipage}%
	\begin{minipage}[htp]{0.5\linewidth}  
		\centering  
		\includegraphics[width=\columnwidth]{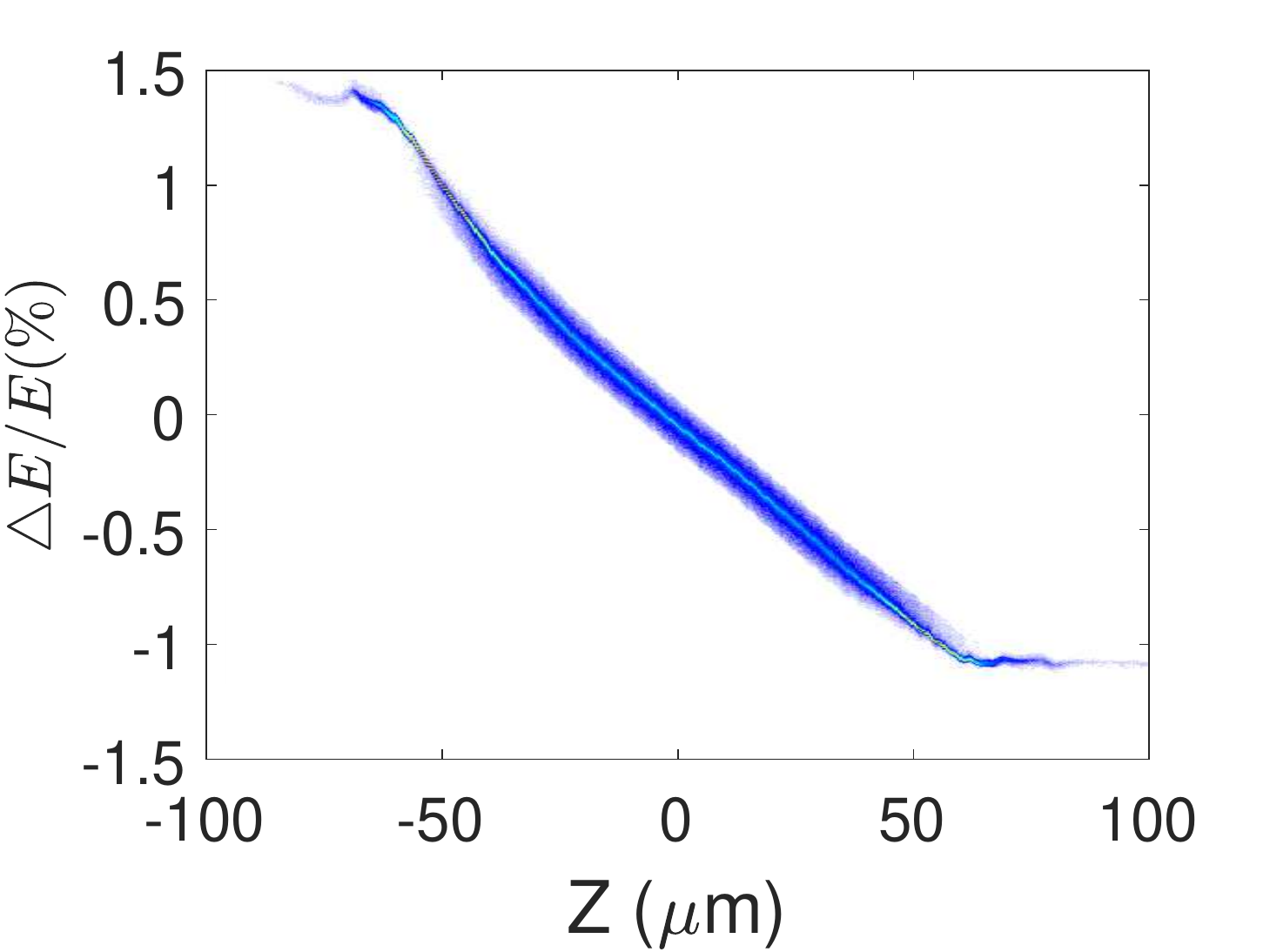}  
	\end{minipage}  
	\begin{minipage}[htp]{0.5\linewidth}  
	\centering  
	\includegraphics[width=\columnwidth]{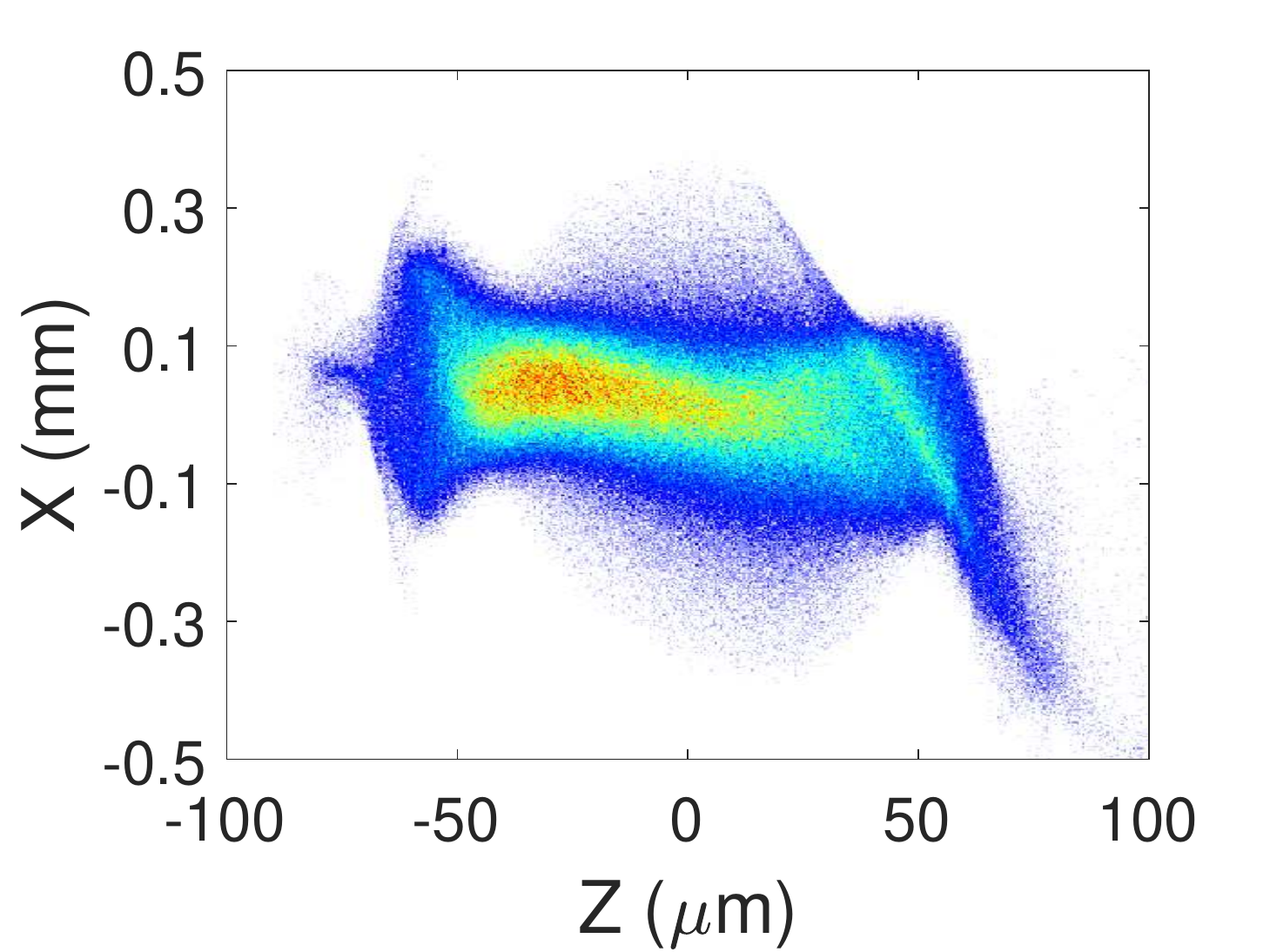}  
\end{minipage}%
\begin{minipage}[htp]{0.5\linewidth}  
	\centering  
	\includegraphics[width=\columnwidth]{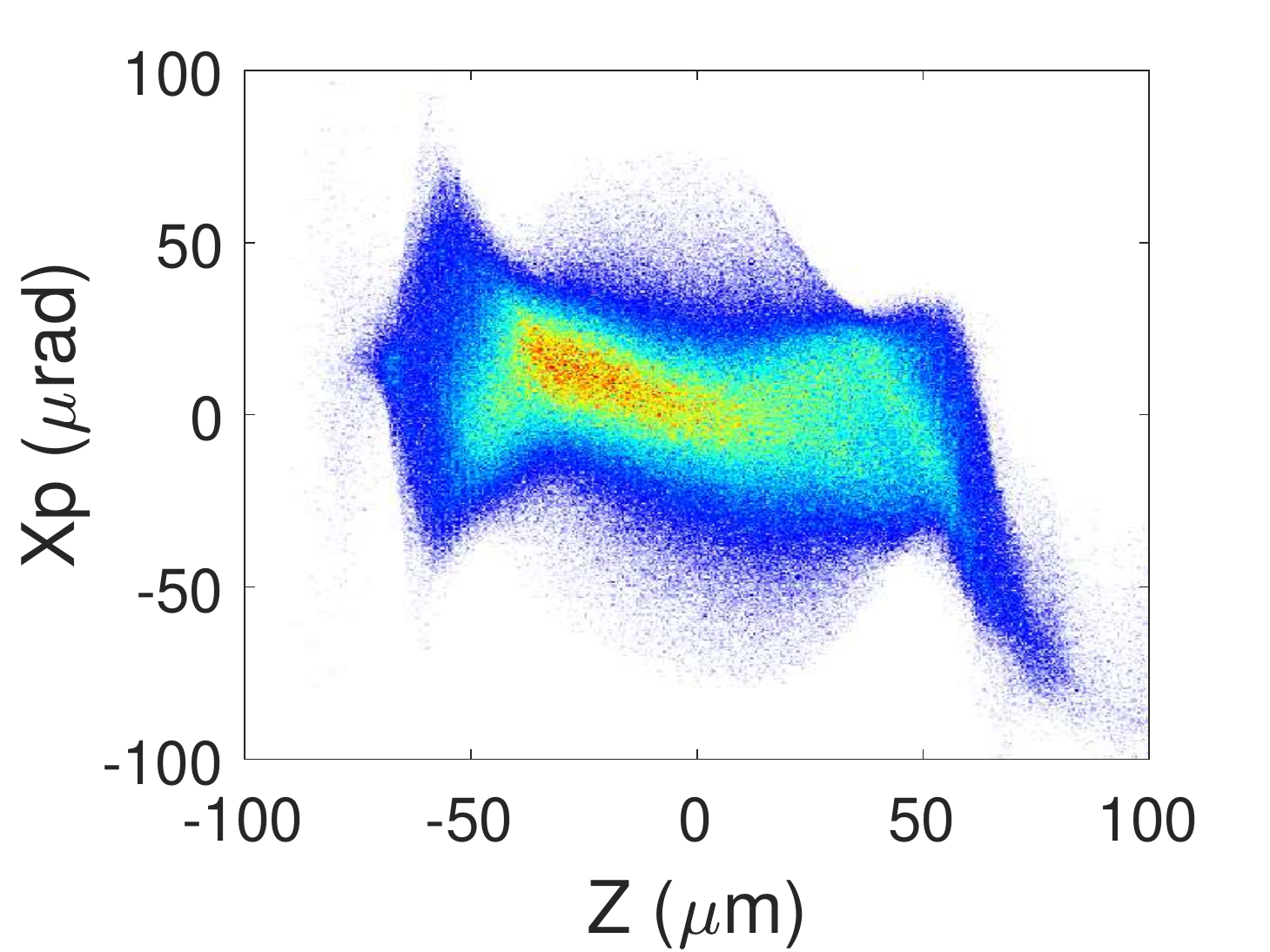}  
\end{minipage}  
	\captionsetup{justification = raggedright, singlelinecheck = false}
	\caption{Current profile (top left), longitudinal phase space (top right) and horizontal phase space (bottom) of the chosen case (see the red dashed line in Fig.\ \ref{FIG5}).}
	\label{FIG6} 
\end{figure} 

\begin{figure}  
	\begin{minipage}[H]{0.5\linewidth}  
		\centering  
		\includegraphics[width=\columnwidth]{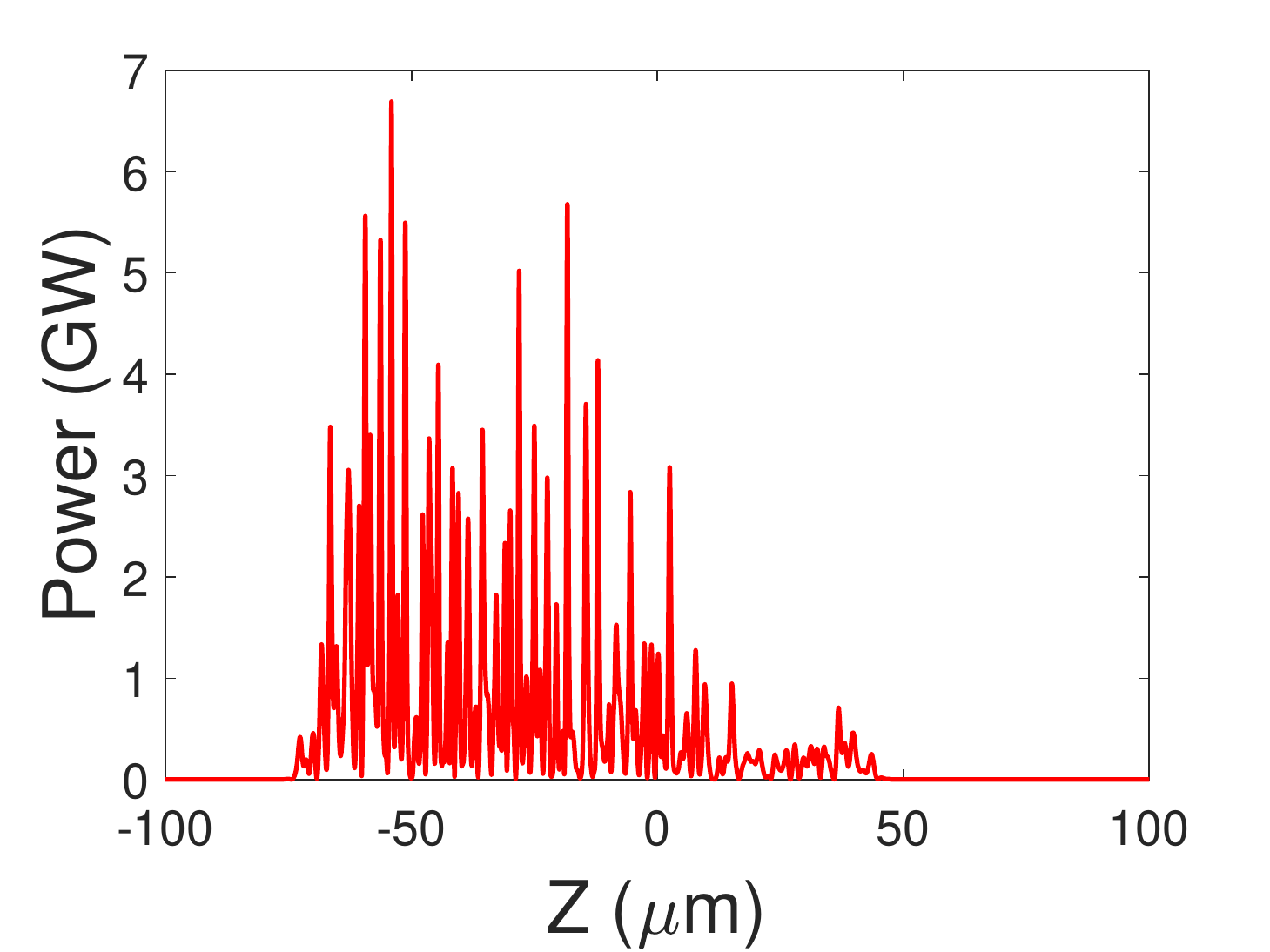}  
	\end{minipage}%
	\begin{minipage}[H]{0.5\linewidth}  
		\centering  
		\includegraphics[width=\columnwidth]{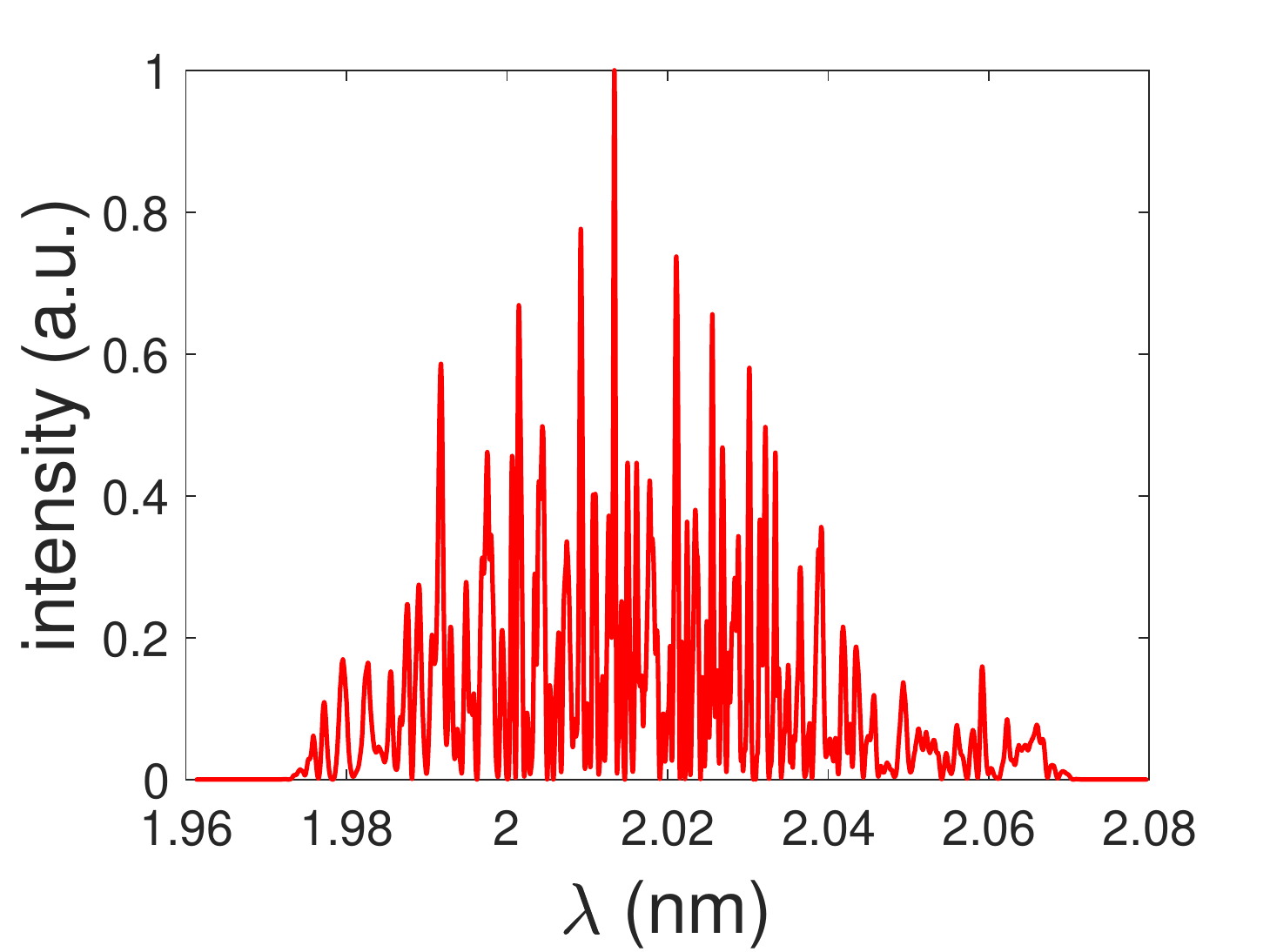}  
	\end{minipage}  

	\caption{Simulated FEL power profile (left) and spectrum (right) of the selected electron bunch. }
	\label{FIG7} 
\end{figure} 

The scheme presented in \cite{25,35} removes the beam yaw by sending energy chirped electron beams into well-controlled dispersion sections. In SXFEL, there are two quadrupoles in the first magnetic chicane that can be used to correct the beam yaw. However, if an optimization result from the Section III is directly corrected by this method, the current profile and energy chirp of this electron bunch may also be changed. In order to get the optimal solutions, here combining the optimization algorithms with the beam yaw correction is considered. Due to the ability of NSGA-III to optimize many objectives, beam qualities affected by over-compression can be optimized simultaneously. In order to take into account the influence of the beam yaw, in addition to the four objectives proposed in Section III, the normalized horizontal emittance at the exit of Linac is chosen as the fifth goal. The smaller projected emittance means the less slice alignment. The strengths of the two quadrupoles in the first bunch compressors are added to optimization variables. In this optimization, only those electron beams with a Gaussian or flat-top shape are pursued. Therefore, besides limiting the peak current and the sign of the energy chirp, a solution with a profile factor less than 70\% will also be given the worst fitness value. Population and iteration in this algorithm are set to 300 and 100.

Pareto front of the last generation is presented in Fig.\ \ref{FIG5} where values of the five objectives are also normalized. In this figure, values of energy chirp, profile factor, peak current, slice energy spread, and normalized horizontal emittance are devided by 2.68\%, 0.94, 2000 A, $6.56\times10^{-4}$, and $75.1\ mm\cdot mrad$ separately. The front shows that the normalized horizontal emittance values of most electron bunches are in a small range, which indicate that the algorithm is practical and efficient for the beam yaw correction. The relationship between the other four objectives is similar to that in Fig.\ \ref{FIG2}.

The maximum energy chirp in this optimization is 2.68\% but the other objectives of the solution with maximum energy chirp do not approach the optimal values. After weighing the five objectives, a solution with an energy chirp of 2.54\% is selected and presented in the Fig.\ \ref{FIG5} by a red dashed line. The peak current, normalized horizontal emittance, and profile factor of the chosen electron bunch are 1330 A, $1.6\ mm\cdot mrad$,  and 0.82. Current distribution and longitudinal phase space of the electron bunch are shown in Fig.\ \ref{FIG6} (top left) and Fig.\ \ref{FIG6} (top right). Fig.\ \ref{FIG6} (bottom) shows the horizontal phase space of the electron bunch in which the corresponding beam yaw has been correctly well. If the two quadrupoles are not used, the final normalized horizontal emittance is $3\ mm\cdot mrad$ and there is a clear beam yaw in the horizontal phase space. GENESIS simulation results of the solution are presented in Fig.\ \ref{FIG7}. The XFEL bandwidth is 4.6\% and the pulse energy is 286 $\mu J$. 

 To analyze the effects of the voltages and timing jitters in the rf structure, 1000 ELEGANT runs are performed with randomized voltages and phases of the first accelerating section and linearizer. Phase jitters of the S-band and X-band are 0.1 and 0.4 deg. Voltage jitters in both sections are 0.04\%. The impacts of these jitters on the optimization targets are calculated. The rms jitters of the peak current, profile factor, and energy chirp are 15.14\%, 2.76\%, and 0.84\%. The jitter analyses indicate that the FEL bandwidth is stable in the over-compression mode and the peak power have some fluctuations due to the peak current jitter which is in an acceptable range.

\section{Conclusions}

In this paper, an evolutionary many-objective optimization algorithm has been applied to optimize the over-compression mode in the linac for producing large-bandwidth XFEL pulses. Benefiting from the ability that can optimize more than three goals simultaneously, objectives including the energy chirp, slice energy spread, peak current, and current profile are all considered in the optimization. In addition, the beam yaw correction has been directly included by adding horizontal projected emittance as an objective. In the case of SXFEL user facility, simulations show that the current profile of the electron bunch has a large impact on the maximum available energy chirp. A maximum energy chirp of 4.4\% can be obtained when the electron beam is of poor current uniformity. Considering the broadband FEL lasing, i.e., a good current profile and a beam yaw correction included, our results indicate that the electron beam with 2.54\% energy chirp generates a 2nm FEL pulse with full bandwidth of 4.6\%. 

This algorithm can be easily extented to include more optimizations for other parts of the FEL facility such as the injector. And the accuracy of the optimization result largely depends on the definition and calculation method of the optimization goal. Moreover, it is also possible to optimize the problem based on the GENESIS simulation results when the computing power is sufficient.

\section*{Acknowledgments}
The author would like to thank Z. Wang, M. Zhang, B. Liu, D. Wang and Z. Zhao for helpful discussions on beam dynamics and SXFEL projects; J. G. Power for helpful discussions on beam-driven collinear wakefield accelerators. This work was partially supported by the National Natural Science Foundation of China (11775293), the National Key Research and Development Program of China (2016YFA0401900), the Young Elite Scientist Sponsorship Program by CAST (2015QNRC001) and Ten Thousand Talent Program.


\bibliography{mybibfile}

\begin{thebibliography}{50}%
\makeatletter
\providecommand \@ifxundefined [1]{%
 \@ifx{#1\undefined}
}%
\providecommand \@ifnum [1]{%
 \ifnum #1\expandafter \@firstoftwo
 \else \expandafter \@secondoftwo
 \fi
}%
\providecommand \@ifx [1]{%
 \ifx #1\expandafter \@firstoftwo
 \else \expandafter \@secondoftwo
 \fi
}%
\providecommand \natexlab [1]{#1}%
\providecommand \enquote  [1]{``#1''}%
\providecommand \bibnamefont  [1]{#1}%
\providecommand \bibfnamefont [1]{#1}%
\providecommand \citenamefont [1]{#1}%
\providecommand \href@noop [0]{\@secondoftwo}%
\providecommand \href [0]{\begingroup \@sanitize@url \@href}%
\providecommand \@href[1]{\@@startlink{#1}\@@href}%
\providecommand \@@href[1]{\endgroup#1\@@endlink}%
\providecommand \@sanitize@url [0]{\catcode `\\12\catcode `\$12\catcode
  `\&12\catcode `\#12\catcode `\^12\catcode `\_12\catcode `\%12\relax}%
\providecommand \@@startlink[1]{}%
\providecommand \@@endlink[0]{}%
\providecommand \url  [0]{\begingroup\@sanitize@url \@url }%
\providecommand \@url [1]{\endgroup\@href {#1}{\urlprefix }}%
\providecommand \urlprefix  [0]{URL }%
\providecommand \Eprint [0]{\href }%
\providecommand \doibase [0]{http://dx.doi.org/}%
\providecommand \selectlanguage [0]{\@gobble}%
\providecommand \bibinfo  [0]{\@secondoftwo}%
\providecommand \bibfield  [0]{\@secondoftwo}%
\providecommand \translation [1]{[#1]}%
\providecommand \BibitemOpen [0]{}%
\providecommand \bibitemStop [0]{}%
\providecommand \bibitemNoStop [0]{.\EOS\space}%
\providecommand \EOS [0]{\spacefactor3000\relax}%
\providecommand \BibitemShut  [1]{\csname bibitem#1\endcsname}%
\let\auto@bib@innerbib\@empty
\bibitem [{\citenamefont {Barletta}\ \emph {et~al.}(2010)\citenamefont
  {Barletta}, \citenamefont {Bisognano}, \citenamefont {Corlett}, \citenamefont
  {Emma}, \citenamefont {Huang}, \citenamefont {Kim}, \citenamefont {Lindberg},
  \citenamefont {Murphy}, \citenamefont {Neil}, \citenamefont {Nguyen} \emph
  {et~al.}}]{1}%
  \BibitemOpen
  \bibfield  {author} {\bibinfo {author} {\bibfnamefont {W.}~\bibnamefont
  {Barletta}}, \bibinfo {author} {\bibfnamefont {J.}~\bibnamefont {Bisognano}},
  \bibinfo {author} {\bibfnamefont {J.}~\bibnamefont {Corlett}}, \bibinfo
  {author} {\bibfnamefont {P.}~\bibnamefont {Emma}}, \bibinfo {author}
  {\bibfnamefont {Z.}~\bibnamefont {Huang}}, \bibinfo {author} {\bibfnamefont
  {K.-J.}\ \bibnamefont {Kim}}, \bibinfo {author} {\bibfnamefont
  {R.}~\bibnamefont {Lindberg}}, \bibinfo {author} {\bibfnamefont
  {J.}~\bibnamefont {Murphy}}, \bibinfo {author} {\bibfnamefont
  {G.}~\bibnamefont {Neil}}, \bibinfo {author} {\bibfnamefont {D.}~\bibnamefont
  {Nguyen}},  \emph {et~al.},\ }\href@noop {} {\bibfield  {journal} {\bibinfo
  {journal} {Nuclear Instruments and Methods in Physics Research Section A:
  Accelerators, Spectrometers, Detectors and Associated Equipment}\ }\textbf
  {\bibinfo {volume} {618}},\ \bibinfo {pages} {69} (\bibinfo {year}
  {2010})}\BibitemShut {NoStop}%
\bibitem [{\citenamefont {Emma}\ \emph {et~al.}(2010)\citenamefont {Emma},
  \citenamefont {Akre}, \citenamefont {Arthur}, \citenamefont {Bionta},
  \citenamefont {Bostedt}, \citenamefont {Bozek}, \citenamefont {Brachmann},
  \citenamefont {Bucksbaum}, \citenamefont {Coffee}, \citenamefont {Decker}
  \emph {et~al.}}]{2}%
  \BibitemOpen
  \bibfield  {author} {\bibinfo {author} {\bibfnamefont {P.}~\bibnamefont
  {Emma}}, \bibinfo {author} {\bibfnamefont {R.}~\bibnamefont {Akre}}, \bibinfo
  {author} {\bibfnamefont {J.}~\bibnamefont {Arthur}}, \bibinfo {author}
  {\bibfnamefont {R.}~\bibnamefont {Bionta}}, \bibinfo {author} {\bibfnamefont
  {C.}~\bibnamefont {Bostedt}}, \bibinfo {author} {\bibfnamefont
  {J.}~\bibnamefont {Bozek}}, \bibinfo {author} {\bibfnamefont
  {A.}~\bibnamefont {Brachmann}}, \bibinfo {author} {\bibfnamefont
  {P.}~\bibnamefont {Bucksbaum}}, \bibinfo {author} {\bibfnamefont
  {R.}~\bibnamefont {Coffee}}, \bibinfo {author} {\bibfnamefont {F.-J.}\
  \bibnamefont {Decker}},  \emph {et~al.},\ }\href@noop {} {\bibfield
  {journal} {\bibinfo  {journal} {Nature Photonics}\ }\textbf {\bibinfo
  {volume} {4}},\ \bibinfo {pages} {641} (\bibinfo {year} {2010})}\BibitemShut
  {NoStop}%
\bibitem [{\citenamefont {Kondratenko}\ and\ \citenamefont {Saldin}(1980)}]{4}%
  \BibitemOpen
  \bibfield  {author} {\bibinfo {author} {\bibfnamefont {A.}~\bibnamefont
  {Kondratenko}}\ and\ \bibinfo {author} {\bibfnamefont {E.}~\bibnamefont
  {Saldin}},\ }\href@noop {} {\bibfield  {journal} {\bibinfo  {journal} {Part.
  Accel.}\ }\textbf {\bibinfo {volume} {10}},\ \bibinfo {pages} {207} (\bibinfo
  {year} {1980})}\BibitemShut {NoStop}%
\bibitem [{\citenamefont {Ackermann}\ \emph {et~al.}(2007)\citenamefont
  {Ackermann}, \citenamefont {Asova}, \citenamefont {Ayvazyan}, \citenamefont
  {Azima}, \citenamefont {Baboi}, \citenamefont {B{\"a}hr}, \citenamefont
  {Balandin}, \citenamefont {Beutner}, \citenamefont {Brandt}, \citenamefont
  {Bolzmann} \emph {et~al.}}]{5}%
  \BibitemOpen
  \bibfield  {author} {\bibinfo {author} {\bibfnamefont {W.~a.}\ \bibnamefont
  {Ackermann}}, \bibinfo {author} {\bibfnamefont {G.}~\bibnamefont {Asova}},
  \bibinfo {author} {\bibfnamefont {V.}~\bibnamefont {Ayvazyan}}, \bibinfo
  {author} {\bibfnamefont {A.}~\bibnamefont {Azima}}, \bibinfo {author}
  {\bibfnamefont {N.}~\bibnamefont {Baboi}}, \bibinfo {author} {\bibfnamefont
  {J.}~\bibnamefont {B{\"a}hr}}, \bibinfo {author} {\bibfnamefont
  {V.}~\bibnamefont {Balandin}}, \bibinfo {author} {\bibfnamefont
  {B.}~\bibnamefont {Beutner}}, \bibinfo {author} {\bibfnamefont
  {A.}~\bibnamefont {Brandt}}, \bibinfo {author} {\bibfnamefont
  {A.}~\bibnamefont {Bolzmann}},  \emph {et~al.},\ }\href@noop {} {\bibfield
  {journal} {\bibinfo  {journal} {Nature Photonics}\ }\textbf {\bibinfo
  {volume} {1}},\ \bibinfo {pages} {336} (\bibinfo {year} {2007})}\BibitemShut
  {NoStop}%
\bibitem [{\citenamefont {Bonifacio}\ \emph {et~al.}(1984)\citenamefont
  {Bonifacio}, \citenamefont {Pellegrini},\ and\ \citenamefont {Narducci}}]{6}%
  \BibitemOpen
  \bibfield  {author} {\bibinfo {author} {\bibfnamefont {R.}~\bibnamefont
  {Bonifacio}}, \bibinfo {author} {\bibfnamefont {C.}~\bibnamefont
  {Pellegrini}}, \ and\ \bibinfo {author} {\bibfnamefont {L.}~\bibnamefont
  {Narducci}},\ }\href@noop {} {\bibfield  {journal} {\bibinfo  {journal}
  {Optics Communications}\ }\textbf {\bibinfo {volume} {50}},\ \bibinfo {pages}
  {373} (\bibinfo {year} {1984})}\BibitemShut {NoStop}%
\bibitem [{\citenamefont {Amann}\ \emph {et~al.}(2012)\citenamefont {Amann},
  \citenamefont {Berg}, \citenamefont {Blank}, \citenamefont {Decker},
  \citenamefont {Ding}, \citenamefont {Emma}, \citenamefont {Feng},
  \citenamefont {Frisch}, \citenamefont {Fritz}, \citenamefont {Hastings} \emph
  {et~al.}}]{48}%
  \BibitemOpen
  \bibfield  {author} {\bibinfo {author} {\bibfnamefont {J.}~\bibnamefont
  {Amann}}, \bibinfo {author} {\bibfnamefont {W.}~\bibnamefont {Berg}},
  \bibinfo {author} {\bibfnamefont {V.}~\bibnamefont {Blank}}, \bibinfo
  {author} {\bibfnamefont {F.-J.}\ \bibnamefont {Decker}}, \bibinfo {author}
  {\bibfnamefont {Y.}~\bibnamefont {Ding}}, \bibinfo {author} {\bibfnamefont
  {P.}~\bibnamefont {Emma}}, \bibinfo {author} {\bibfnamefont {Y.}~\bibnamefont
  {Feng}}, \bibinfo {author} {\bibfnamefont {J.}~\bibnamefont {Frisch}},
  \bibinfo {author} {\bibfnamefont {D.}~\bibnamefont {Fritz}}, \bibinfo
  {author} {\bibfnamefont {J.}~\bibnamefont {Hastings}},  \emph {et~al.},\
  }\href@noop {} {\bibfield  {journal} {\bibinfo  {journal} {Nature Photonics}\
  }\textbf {\bibinfo {volume} {6}},\ \bibinfo {pages} {693} (\bibinfo {year}
  {2012})}\BibitemShut {NoStop}%
\bibitem [{\citenamefont {Allaria}\ \emph {et~al.}(2012)\citenamefont
  {Allaria}, \citenamefont {Appio}, \citenamefont {Badano}, \citenamefont
  {Barletta}, \citenamefont {Bassanese}, \citenamefont {Biedron}, \citenamefont
  {Borga}, \citenamefont {Busetto}, \citenamefont {Castronovo}, \citenamefont
  {Cinquegrana} \emph {et~al.}}]{49}%
  \BibitemOpen
  \bibfield  {author} {\bibinfo {author} {\bibfnamefont {E.}~\bibnamefont
  {Allaria}}, \bibinfo {author} {\bibfnamefont {R.}~\bibnamefont {Appio}},
  \bibinfo {author} {\bibfnamefont {L.}~\bibnamefont {Badano}}, \bibinfo
  {author} {\bibfnamefont {W.}~\bibnamefont {Barletta}}, \bibinfo {author}
  {\bibfnamefont {S.}~\bibnamefont {Bassanese}}, \bibinfo {author}
  {\bibfnamefont {S.}~\bibnamefont {Biedron}}, \bibinfo {author} {\bibfnamefont
  {A.}~\bibnamefont {Borga}}, \bibinfo {author} {\bibfnamefont
  {E.}~\bibnamefont {Busetto}}, \bibinfo {author} {\bibfnamefont
  {D.}~\bibnamefont {Castronovo}}, \bibinfo {author} {\bibfnamefont
  {P.}~\bibnamefont {Cinquegrana}},  \emph {et~al.},\ }\href@noop {} {\bibfield
   {journal} {\bibinfo  {journal} {Nature Photonics}\ }\textbf {\bibinfo
  {volume} {6}},\ \bibinfo {pages} {699} (\bibinfo {year} {2012})}\BibitemShut
  {NoStop}%
\bibitem [{\citenamefont {Deng}\ and\ \citenamefont {Feng}(2013)}]{7}%
  \BibitemOpen
  \bibfield  {author} {\bibinfo {author} {\bibfnamefont {H.}~\bibnamefont
  {Deng}}\ and\ \bibinfo {author} {\bibfnamefont {C.}~\bibnamefont {Feng}},\
  }\href@noop {} {\bibfield  {journal} {\bibinfo  {journal} {Physical Review
  Letters}\ }\textbf {\bibinfo {volume} {111}},\ \bibinfo {pages} {084801}
  (\bibinfo {year} {2013})}\BibitemShut {NoStop}%
\bibitem [{\citenamefont {Stupakov}(2009)}]{8}%
  \BibitemOpen
  \bibfield  {author} {\bibinfo {author} {\bibfnamefont {G.}~\bibnamefont
  {Stupakov}},\ }\href@noop {} {\bibfield  {journal} {\bibinfo  {journal}
  {Physical Review Letters}\ }\textbf {\bibinfo {volume} {102}},\ \bibinfo
  {pages} {074801} (\bibinfo {year} {2009})}\BibitemShut {NoStop}%
\bibitem [{\citenamefont {Yu}(1991)}]{9}%
  \BibitemOpen
  \bibfield  {author} {\bibinfo {author} {\bibfnamefont {L.~H.}\ \bibnamefont
  {Yu}},\ }\href@noop {} {\bibfield  {journal} {\bibinfo  {journal} {Physical
  Review A}\ }\textbf {\bibinfo {volume} {44}},\ \bibinfo {pages} {5178}
  (\bibinfo {year} {1991})}\BibitemShut {NoStop}%
\bibitem [{\citenamefont {McNeil}\ \emph {et~al.}(2013)\citenamefont {McNeil},
  \citenamefont {Thompson},\ and\ \citenamefont {Dunning}}]{50}%
  \BibitemOpen
  \bibfield  {author} {\bibinfo {author} {\bibfnamefont {B.}~\bibnamefont
  {McNeil}}, \bibinfo {author} {\bibfnamefont {N.}~\bibnamefont {Thompson}}, \
  and\ \bibinfo {author} {\bibfnamefont {D.}~\bibnamefont {Dunning}},\
  }\href@noop {} {\bibfield  {journal} {\bibinfo  {journal} {Physical Review
  Letters}\ }\textbf {\bibinfo {volume} {110}},\ \bibinfo {pages} {134802}
  (\bibinfo {year} {2013})}\BibitemShut {NoStop}%
\bibitem [{\citenamefont {Dai}\ \emph {et~al.}(2012)\citenamefont {Dai},
  \citenamefont {Deng},\ and\ \citenamefont {Dai}}]{51}%
  \BibitemOpen
  \bibfield  {author} {\bibinfo {author} {\bibfnamefont {J.}~\bibnamefont
  {Dai}}, \bibinfo {author} {\bibfnamefont {H.}~\bibnamefont {Deng}}, \ and\
  \bibinfo {author} {\bibfnamefont {Z.}~\bibnamefont {Dai}},\ }\href@noop {}
  {\bibfield  {journal} {\bibinfo  {journal} {Physical Review Letters}\
  }\textbf {\bibinfo {volume} {108}},\ \bibinfo {pages} {034802} (\bibinfo
  {year} {2012})}\BibitemShut {NoStop}%
\bibitem [{\citenamefont {Li}\ \emph {et~al.}(2018)\citenamefont {Li},
  \citenamefont {Yan}, \citenamefont {Feng}, \citenamefont {Zhang},\ and\
  \citenamefont {Deng}}]{52}%
  \BibitemOpen
  \bibfield  {author} {\bibinfo {author} {\bibfnamefont {K.}~\bibnamefont
  {Li}}, \bibinfo {author} {\bibfnamefont {J.}~\bibnamefont {Yan}}, \bibinfo
  {author} {\bibfnamefont {C.}~\bibnamefont {Feng}}, \bibinfo {author}
  {\bibfnamefont {M.}~\bibnamefont {Zhang}}, \ and\ \bibinfo {author}
  {\bibfnamefont {H.}~\bibnamefont {Deng}},\ }\href@noop {} {\bibfield
  {journal} {\bibinfo  {journal} {Physical Review Accelerators and Beams}\
  }\textbf {\bibinfo {volume} {21}},\ \bibinfo {pages} {040702} (\bibinfo
  {year} {2018})}\BibitemShut {NoStop}%
\bibitem [{\citenamefont {Ganter}\ and\ \citenamefont {ed}(2010)}]{10}%
  \BibitemOpen
  \bibfield  {author} {\bibinfo {author} {\bibfnamefont {R.}~\bibnamefont
  {Ganter}}\ and\ \bibinfo {author} {\bibnamefont {ed}},\ }\href@noop {}
  {\bibfield  {journal} {\bibinfo  {journal} {Appeal Rev.current L. and
  L.reform}\ } (\bibinfo {year} {2010})}\BibitemShut {NoStop}%
\bibitem [{\citenamefont {Baradaran}\ \emph
  {et~al.}(2013{\natexlab{a}})\citenamefont {Baradaran}, \citenamefont
  {Berrisford}, \citenamefont {Minhas},\ and\ \citenamefont {Sazanov}}]{11}%
  \BibitemOpen
  \bibfield  {author} {\bibinfo {author} {\bibfnamefont {R.}~\bibnamefont
  {Baradaran}}, \bibinfo {author} {\bibfnamefont {J.~M.}\ \bibnamefont
  {Berrisford}}, \bibinfo {author} {\bibfnamefont {G.~S.}\ \bibnamefont
  {Minhas}}, \ and\ \bibinfo {author} {\bibfnamefont {L.~A.}\ \bibnamefont
  {Sazanov}},\ }\href@noop {} {\bibfield  {journal} {\bibinfo  {journal}
  {Nature}\ }\textbf {\bibinfo {volume} {494}},\ \bibinfo {pages} {443}
  (\bibinfo {year} {2013}{\natexlab{a}})}\BibitemShut {NoStop}%
\bibitem [{\citenamefont {Patterson}\ \emph {et~al.}(2010)\citenamefont
  {Patterson}, \citenamefont {Abela}, \citenamefont {Braun}, \citenamefont
  {Flechsig}, \citenamefont {Ganter}, \citenamefont {Kim}, \citenamefont
  {Kirk}, \citenamefont {Oppelt}, \citenamefont {Pedrozzi},\ and\ \citenamefont
  {Reiche}}]{12}%
  \BibitemOpen
  \bibfield  {author} {\bibinfo {author} {\bibfnamefont {B.~D.}\ \bibnamefont
  {Patterson}}, \bibinfo {author} {\bibfnamefont {R.}~\bibnamefont {Abela}},
  \bibinfo {author} {\bibfnamefont {H.~H.}\ \bibnamefont {Braun}}, \bibinfo
  {author} {\bibfnamefont {U.}~\bibnamefont {Flechsig}}, \bibinfo {author}
  {\bibfnamefont {R.}~\bibnamefont {Ganter}}, \bibinfo {author} {\bibfnamefont
  {Y.}~\bibnamefont {Kim}}, \bibinfo {author} {\bibfnamefont {E.}~\bibnamefont
  {Kirk}}, \bibinfo {author} {\bibfnamefont {A.}~\bibnamefont {Oppelt}},
  \bibinfo {author} {\bibfnamefont {M.}~\bibnamefont {Pedrozzi}}, \ and\
  \bibinfo {author} {\bibfnamefont {S.}~\bibnamefont {Reiche}},\ }\href@noop {}
  {\bibfield  {journal} {\bibinfo  {journal} {New Journal of Physics}\ }\textbf
  {\bibinfo {volume} {12}},\ \bibinfo {pages} {035012} (\bibinfo {year}
  {2010})}\BibitemShut {NoStop}%
\bibitem [{\citenamefont {Hendrickson}\ and\ \citenamefont {Ogata}(1997)}]{13}%
  \BibitemOpen
  \bibfield  {author} {\bibinfo {author} {\bibfnamefont {W.~A.}\ \bibnamefont
  {Hendrickson}}\ and\ \bibinfo {author} {\bibfnamefont {C.~M.}\ \bibnamefont
  {Ogata}},\ }\href@noop {} {\bibfield  {journal} {\bibinfo  {journal} {Methods
  in Enzymology}\ }\textbf {\bibinfo {volume} {276}},\ \bibinfo {pages} {494}
  (\bibinfo {year} {1997})}\BibitemShut {NoStop}%
\bibitem [{\citenamefont {Son}\ \emph {et~al.}(2011)\citenamefont {Son},
  \citenamefont {Chapman}, \citenamefont {Santra} \emph {et~al.}}]{14}%
  \BibitemOpen
  \bibfield  {author} {\bibinfo {author} {\bibfnamefont {S.-K.}\ \bibnamefont
  {Son}}, \bibinfo {author} {\bibfnamefont {H.~N.}\ \bibnamefont {Chapman}},
  \bibinfo {author} {\bibfnamefont {R.}~\bibnamefont {Santra}},  \emph
  {et~al.},\ }\href@noop {} {\bibfield  {journal} {\bibinfo  {journal}
  {Physical Review Letters}\ }\textbf {\bibinfo {volume} {107}},\ \bibinfo
  {pages} {218102} (\bibinfo {year} {2011})}\BibitemShut {NoStop}%
\bibitem [{\citenamefont {Baradaran}\ \emph
  {et~al.}(2013{\natexlab{b}})\citenamefont {Baradaran}, \citenamefont
  {Berrisford}, \citenamefont {Minhas},\ and\ \citenamefont {Sazanov}}]{15}%
  \BibitemOpen
  \bibfield  {author} {\bibinfo {author} {\bibfnamefont {R.}~\bibnamefont
  {Baradaran}}, \bibinfo {author} {\bibfnamefont {J.~M.}\ \bibnamefont
  {Berrisford}}, \bibinfo {author} {\bibfnamefont {G.~S.}\ \bibnamefont
  {Minhas}}, \ and\ \bibinfo {author} {\bibfnamefont {L.~A.}\ \bibnamefont
  {Sazanov}},\ }\href@noop {} {\bibfield  {journal} {\bibinfo  {journal}
  {Nature}\ }\textbf {\bibinfo {volume} {494}},\ \bibinfo {pages} {443}
  (\bibinfo {year} {2013}{\natexlab{b}})}\BibitemShut {NoStop}%
\bibitem [{\citenamefont {Dejoie}\ \emph {et~al.}(2015)\citenamefont {Dejoie},
  \citenamefont {Smeets}, \citenamefont {Baerlocher}, \citenamefont {Tamura},
  \citenamefont {Pattison}, \citenamefont {Abela},\ and\ \citenamefont
  {McCusker}}]{16}%
  \BibitemOpen
  \bibfield  {author} {\bibinfo {author} {\bibfnamefont {C.}~\bibnamefont
  {Dejoie}}, \bibinfo {author} {\bibfnamefont {S.}~\bibnamefont {Smeets}},
  \bibinfo {author} {\bibfnamefont {C.}~\bibnamefont {Baerlocher}}, \bibinfo
  {author} {\bibfnamefont {N.}~\bibnamefont {Tamura}}, \bibinfo {author}
  {\bibfnamefont {P.}~\bibnamefont {Pattison}}, \bibinfo {author}
  {\bibfnamefont {R.}~\bibnamefont {Abela}}, \ and\ \bibinfo {author}
  {\bibfnamefont {L.~B.}\ \bibnamefont {McCusker}},\ }\href@noop {} {\bibfield
  {journal} {\bibinfo  {journal} {IUCrJ}\ }\textbf {\bibinfo {volume} {2}},\
  \bibinfo {pages} {361} (\bibinfo {year} {2015})}\BibitemShut {NoStop}%
\bibitem [{\citenamefont {Huang}\ and\ \citenamefont {Kim}(2007)}]{17}%
  \BibitemOpen
  \bibfield  {author} {\bibinfo {author} {\bibfnamefont {Z.}~\bibnamefont
  {Huang}}\ and\ \bibinfo {author} {\bibfnamefont {K.-J.}\ \bibnamefont
  {Kim}},\ }\href@noop {} {\bibfield  {journal} {\bibinfo  {journal} {Phys.
  Rev. ST Accel. Beams}\ }\textbf {\bibinfo {volume} {10}},\ \bibinfo {pages}
  {349} (\bibinfo {year} {2007})}\BibitemShut {NoStop}%
\bibitem [{\citenamefont {Prat}\ \emph {et~al.}(2016)\citenamefont {Prat},
  \citenamefont {Calvi},\ and\ \citenamefont {Reiche}}]{22}%
  \BibitemOpen
  \bibfield  {author} {\bibinfo {author} {\bibfnamefont {E.}~\bibnamefont
  {Prat}}, \bibinfo {author} {\bibfnamefont {M.}~\bibnamefont {Calvi}}, \ and\
  \bibinfo {author} {\bibfnamefont {S.}~\bibnamefont {Reiche}},\ }\href@noop {}
  {\bibfield  {journal} {\bibinfo  {journal} {Journal of Synchrotron
  Radiation}\ }\textbf {\bibinfo {volume} {23}},\ \bibinfo {pages} {874}
  (\bibinfo {year} {2016})}\BibitemShut {NoStop}%
\bibitem [{\citenamefont {Song}\ \emph {et~al.}(2018)\citenamefont {Song},
  \citenamefont {Yan}, \citenamefont {Li}, \citenamefont {Feng},\ and\
  \citenamefont {Deng}}]{23}%
  \BibitemOpen
  \bibfield  {author} {\bibinfo {author} {\bibfnamefont {M.}~\bibnamefont
  {Song}}, \bibinfo {author} {\bibfnamefont {J.}~\bibnamefont {Yan}}, \bibinfo
  {author} {\bibfnamefont {K.}~\bibnamefont {Li}}, \bibinfo {author}
  {\bibfnamefont {C.}~\bibnamefont {Feng}}, \ and\ \bibinfo {author}
  {\bibfnamefont {H.}~\bibnamefont {Deng}},\ }\href@noop {} {\bibfield
  {journal} {\bibinfo  {journal} {Nuclear Instruments and Methods in Physics
  Research}\ }\textbf {\bibinfo {volume} {884}},\ \bibinfo {pages} {11}
  (\bibinfo {year} {2018})}\BibitemShut {NoStop}%
\bibitem [{\citenamefont {Hernandez}\ \emph {et~al.}(2016)\citenamefont
  {Hernandez}, \citenamefont {Prat}, \citenamefont {Bettoni}, \citenamefont
  {Beutner},\ and\ \citenamefont {Reiche}}]{19}%
  \BibitemOpen
  \bibfield  {author} {\bibinfo {author} {\bibfnamefont {A.~S.}\ \bibnamefont
  {Hernandez}}, \bibinfo {author} {\bibfnamefont {E.}~\bibnamefont {Prat}},
  \bibinfo {author} {\bibfnamefont {S.}~\bibnamefont {Bettoni}}, \bibinfo
  {author} {\bibfnamefont {B.}~\bibnamefont {Beutner}}, \ and\ \bibinfo
  {author} {\bibfnamefont {S.}~\bibnamefont {Reiche}},\ }\href@noop {}
  {\bibfield  {journal} {\bibinfo  {journal} {Phys. Rev. ST Accel. Beams}\
  }\textbf {\bibinfo {volume} {19}} (\bibinfo {year} {2016})}\BibitemShut
  {NoStop}%
\bibitem [{\citenamefont {Serkez}\ \emph {et~al.}(2013)\citenamefont {Serkez},
  \citenamefont {Kocharyan}, \citenamefont {Saldin}, \citenamefont
  {Zagorodnov}, \citenamefont {Geloni},\ and\ \citenamefont {Yefanov}}]{20}%
  \BibitemOpen
  \bibfield  {author} {\bibinfo {author} {\bibfnamefont {S.}~\bibnamefont
  {Serkez}}, \bibinfo {author} {\bibfnamefont {V.}~\bibnamefont {Kocharyan}},
  \bibinfo {author} {\bibfnamefont {E.}~\bibnamefont {Saldin}}, \bibinfo
  {author} {\bibfnamefont {I.}~\bibnamefont {Zagorodnov}}, \bibinfo {author}
  {\bibfnamefont {G.}~\bibnamefont {Geloni}}, \ and\ \bibinfo {author}
  {\bibfnamefont {O.}~\bibnamefont {Yefanov}},\ }\href@noop {} {\bibfield
  {journal} {\bibinfo  {journal} {Proceedings of Fel New York Ny Usa}\ }\textbf
  {\bibinfo {volume} {2}},\ \bibinfo {pages} {345} (\bibinfo {year}
  {2013})}\BibitemShut {NoStop}%
\bibitem [{\citenamefont {Zagorodnov}\ \emph {et~al.}(2016)\citenamefont
  {Zagorodnov}, \citenamefont {Feng},\ and\ \citenamefont {Limberg}}]{21}%
  \BibitemOpen
  \bibfield  {author} {\bibinfo {author} {\bibfnamefont {I.}~\bibnamefont
  {Zagorodnov}}, \bibinfo {author} {\bibfnamefont {G.}~\bibnamefont {Feng}}, \
  and\ \bibinfo {author} {\bibfnamefont {T.}~\bibnamefont {Limberg}},\
  }\href@noop {} {\bibfield  {journal} {\bibinfo  {journal} {Nuclear
  Instruments and Methods in Physics Research Section A: Accelerators,
  Spectrometers, Detectors and Associated Equipment}\ }\textbf {\bibinfo
  {volume} {837}},\ \bibinfo {pages} {69} (\bibinfo {year} {2016})}\BibitemShut
  {NoStop}%
\bibitem [{\citenamefont {DEB}(2002)}]{27}%
  \BibitemOpen
  \bibfield  {author} {\bibinfo {author} {\bibfnamefont {K.}~\bibnamefont
  {DEB}},\ }\href@noop {} {\bibfield  {journal} {\bibinfo  {journal} {IEEE
  Trans. Evol. Comput.}\ }\textbf {\bibinfo {volume} {6}} (\bibinfo {year}
  {2002})}\BibitemShut {NoStop}%
\bibitem [{\citenamefont {Bazarov}\ and\ \citenamefont {Sinclair}(2005)}]{28}%
  \BibitemOpen
  \bibfield  {author} {\bibinfo {author} {\bibfnamefont {I.~V.}\ \bibnamefont
  {Bazarov}}\ and\ \bibinfo {author} {\bibfnamefont {C.~K.}\ \bibnamefont
  {Sinclair}},\ }\href@noop {} {\bibfield  {journal} {\bibinfo  {journal}
  {Review of Modern Physics}\ }\textbf {\bibinfo {volume} {8}},\  (\bibinfo
  {year} {2005})}\BibitemShut {NoStop}%
\bibitem [{\citenamefont {Melzak}\ \emph {et~al.}(2012)\citenamefont {Melzak},
  \citenamefont {Mateescu}, \citenamefont {Toca-Herrera},\ and\ \citenamefont
  {Jonas}}]{29}%
  \BibitemOpen
  \bibfield  {author} {\bibinfo {author} {\bibfnamefont {K.~A.}\ \bibnamefont
  {Melzak}}, \bibinfo {author} {\bibfnamefont {A.}~\bibnamefont {Mateescu}},
  \bibinfo {author} {\bibfnamefont {J.~L.}\ \bibnamefont {Toca-Herrera}}, \
  and\ \bibinfo {author} {\bibfnamefont {U.}~\bibnamefont {Jonas}},\
  }\href@noop {} {\bibfield  {journal} {\bibinfo  {journal} {Physical Review
  Special Topics - Accelerators and Beams}\ }\textbf {\bibinfo {volume} {15}},\
  \bibinfo {pages} {49} (\bibinfo {year} {2012})}\BibitemShut {NoStop}%
\bibitem [{\citenamefont {Gao}\ \emph {et~al.}(2011)\citenamefont {Gao},
  \citenamefont {Wang},\ and\ \citenamefont {Li}}]{30}%
  \BibitemOpen
  \bibfield  {author} {\bibinfo {author} {\bibfnamefont {W.}~\bibnamefont
  {Gao}}, \bibinfo {author} {\bibfnamefont {L.}~\bibnamefont {Wang}}, \ and\
  \bibinfo {author} {\bibfnamefont {W.}~\bibnamefont {Li}},\ }\href@noop {}
  {\bibfield  {journal} {\bibinfo  {journal} {Physical Review Special Topics -
  Accelerators and Beams}\ }\textbf {\bibinfo {volume} {14}},\ \bibinfo {pages}
  {2149} (\bibinfo {year} {2011})}\BibitemShut {NoStop}%
\bibitem [{\citenamefont {Wu}\ \emph {et~al.}(2017)\citenamefont {Wu},
  \citenamefont {Hu}, \citenamefont {Setiawan}, \citenamefont {Huang},
  \citenamefont {Raubenheimer}, \citenamefont {Jiao}, \citenamefont {Yu},
  \citenamefont {Mandlekar}, \citenamefont {Spampinati}, \citenamefont {Fang}
  \emph {et~al.}}]{31}%
  \BibitemOpen
  \bibfield  {author} {\bibinfo {author} {\bibfnamefont {J.}~\bibnamefont
  {Wu}}, \bibinfo {author} {\bibfnamefont {N.}~\bibnamefont {Hu}}, \bibinfo
  {author} {\bibfnamefont {H.}~\bibnamefont {Setiawan}}, \bibinfo {author}
  {\bibfnamefont {X.}~\bibnamefont {Huang}}, \bibinfo {author} {\bibfnamefont
  {T.~O.}\ \bibnamefont {Raubenheimer}}, \bibinfo {author} {\bibfnamefont
  {Y.}~\bibnamefont {Jiao}}, \bibinfo {author} {\bibfnamefont {G.}~\bibnamefont
  {Yu}}, \bibinfo {author} {\bibfnamefont {A.}~\bibnamefont {Mandlekar}},
  \bibinfo {author} {\bibfnamefont {S.}~\bibnamefont {Spampinati}}, \bibinfo
  {author} {\bibfnamefont {K.}~\bibnamefont {Fang}},  \emph {et~al.},\
  }\href@noop {} {\bibfield  {journal} {\bibinfo  {journal} {Nuclear
  Instruments and Methods in Physics Research Section A: Accelerators,
  Spectrometers, Detectors and Associated Equipment}\ }\textbf {\bibinfo
  {volume} {846}},\ \bibinfo {pages} {56} (\bibinfo {year} {2017})}\BibitemShut
  {NoStop}%
\bibitem [{\citenamefont {Hofler}\ \emph {et~al.}(2013)\citenamefont {Hofler},
  \citenamefont {Terzi{\'c}}, \citenamefont {Kramer}, \citenamefont {Zvezdin},
  \citenamefont {Morozov}, \citenamefont {Roblin}, \citenamefont {Lin},\ and\
  \citenamefont {Jarvis}}]{32}%
  \BibitemOpen
  \bibfield  {author} {\bibinfo {author} {\bibfnamefont {A.}~\bibnamefont
  {Hofler}}, \bibinfo {author} {\bibfnamefont {B.}~\bibnamefont {Terzi{\'c}}},
  \bibinfo {author} {\bibfnamefont {M.}~\bibnamefont {Kramer}}, \bibinfo
  {author} {\bibfnamefont {A.}~\bibnamefont {Zvezdin}}, \bibinfo {author}
  {\bibfnamefont {V.}~\bibnamefont {Morozov}}, \bibinfo {author} {\bibfnamefont
  {Y.}~\bibnamefont {Roblin}}, \bibinfo {author} {\bibfnamefont
  {F.}~\bibnamefont {Lin}}, \ and\ \bibinfo {author} {\bibfnamefont
  {C.}~\bibnamefont {Jarvis}},\ }\href@noop {} {\bibfield  {journal} {\bibinfo
  {journal} {Physical Review Special Topics-Accelerators and Beams}\ }\textbf
  {\bibinfo {volume} {16}},\ \bibinfo {pages} {010101} (\bibinfo {year}
  {2013})}\BibitemShut {NoStop}%
\bibitem [{\citenamefont {Ikeda}\ \emph {et~al.}(2001)\citenamefont {Ikeda},
  \citenamefont {Kita},\ and\ \citenamefont {Kobayashi}}]{33}%
  \BibitemOpen
  \bibfield  {author} {\bibinfo {author} {\bibfnamefont {K.}~\bibnamefont
  {Ikeda}}, \bibinfo {author} {\bibfnamefont {H.}~\bibnamefont {Kita}}, \ and\
  \bibinfo {author} {\bibfnamefont {S.}~\bibnamefont {Kobayashi}},\ }in\
  \href@noop {} {\emph {\bibinfo {booktitle} {Evolutionary Computation, 2001.
  Proceedings of the 2001 Congress on}}},\ Vol.~\bibinfo {volume} {2}\
  (\bibinfo {organization} {IEEE},\ \bibinfo {year} {2001})\ pp.\ \bibinfo
  {pages} {957--962}\BibitemShut {NoStop}%
\bibitem [{\citenamefont {Deb}\ and\ \citenamefont {Jain}(2014)}]{34}%
  \BibitemOpen
  \bibfield  {author} {\bibinfo {author} {\bibfnamefont {K.}~\bibnamefont
  {Deb}}\ and\ \bibinfo {author} {\bibfnamefont {H.}~\bibnamefont {Jain}},\
  }\href@noop {} {\bibfield  {journal} {\bibinfo  {journal} {IEEE Transactions
  on Evolutionary Computation}\ }\textbf {\bibinfo {volume} {18}},\ \bibinfo
  {pages} {577} (\bibinfo {year} {2014})}\BibitemShut {NoStop}%
\bibitem [{\citenamefont {Jain}\ and\ \citenamefont {Deb}(2014)}]{42}%
  \BibitemOpen
  \bibfield  {author} {\bibinfo {author} {\bibfnamefont {H.}~\bibnamefont
  {Jain}}\ and\ \bibinfo {author} {\bibfnamefont {K.}~\bibnamefont {Deb}},\
  }\href@noop {} {\bibfield  {journal} {\bibinfo  {journal} {IEEE Trans.
  Evolutionary Computation}\ }\textbf {\bibinfo {volume} {18}},\ \bibinfo
  {pages} {602} (\bibinfo {year} {2014})}\BibitemShut {NoStop}%
\bibitem [{\citenamefont {Mkaouer}\ \emph {et~al.}(2014)\citenamefont
  {Mkaouer}, \citenamefont {Kessentini}, \citenamefont {Bechikh}, \citenamefont
  {Deb},\ and\ \citenamefont {{\'O}~Cinn{\'e}ide}}]{44}%
  \BibitemOpen
  \bibfield  {author} {\bibinfo {author} {\bibfnamefont {M.~W.}\ \bibnamefont
  {Mkaouer}}, \bibinfo {author} {\bibfnamefont {M.}~\bibnamefont {Kessentini}},
  \bibinfo {author} {\bibfnamefont {S.}~\bibnamefont {Bechikh}}, \bibinfo
  {author} {\bibfnamefont {K.}~\bibnamefont {Deb}}, \ and\ \bibinfo {author}
  {\bibfnamefont {M.}~\bibnamefont {{\'O}~Cinn{\'e}ide}},\ }in\ \href@noop {}
  {\emph {\bibinfo {booktitle} {Proceedings of the 2014 Annual Conference on
  Genetic and Evolutionary Computation}}}\ (\bibinfo {organization} {ACM},\
  \bibinfo {year} {2014})\ pp.\ \bibinfo {pages} {1263--1270}\BibitemShut
  {NoStop}%
\bibitem [{\citenamefont {Tavana}\ \emph {et~al.}(2016)\citenamefont {Tavana},
  \citenamefont {Li}, \citenamefont {Mobin}, \citenamefont {Komaki},\ and\
  \citenamefont {Teymourian}}]{45}%
  \BibitemOpen
  \bibfield  {author} {\bibinfo {author} {\bibfnamefont {M.}~\bibnamefont
  {Tavana}}, \bibinfo {author} {\bibfnamefont {Z.}~\bibnamefont {Li}}, \bibinfo
  {author} {\bibfnamefont {M.}~\bibnamefont {Mobin}}, \bibinfo {author}
  {\bibfnamefont {M.}~\bibnamefont {Komaki}}, \ and\ \bibinfo {author}
  {\bibfnamefont {E.}~\bibnamefont {Teymourian}},\ }\href@noop {} {\bibfield
  {journal} {\bibinfo  {journal} {Expert Systems with Applications}\ }\textbf
  {\bibinfo {volume} {50}},\ \bibinfo {pages} {17} (\bibinfo {year}
  {2016})}\BibitemShut {NoStop}%
\bibitem [{\citenamefont {Yuan}\ \emph {et~al.}(2015)\citenamefont {Yuan},
  \citenamefont {Tian}, \citenamefont {Yuan}, \citenamefont {Huang},\ and\
  \citenamefont {Ikram}}]{46}%
  \BibitemOpen
  \bibfield  {author} {\bibinfo {author} {\bibfnamefont {X.}~\bibnamefont
  {Yuan}}, \bibinfo {author} {\bibfnamefont {H.}~\bibnamefont {Tian}}, \bibinfo
  {author} {\bibfnamefont {Y.}~\bibnamefont {Yuan}}, \bibinfo {author}
  {\bibfnamefont {Y.}~\bibnamefont {Huang}}, \ and\ \bibinfo {author}
  {\bibfnamefont {R.~M.}\ \bibnamefont {Ikram}},\ }\href@noop {} {\bibfield
  {journal} {\bibinfo  {journal} {Energy Conversion and Management}\ }\textbf
  {\bibinfo {volume} {96}},\ \bibinfo {pages} {568} (\bibinfo {year}
  {2015})}\BibitemShut {NoStop}%
\bibitem [{\citenamefont {Saldin}\ \emph {et~al.}(1997)\citenamefont {Saldin},
  \citenamefont {Schneidmiller},\ and\ \citenamefont {Yurkov}}]{47}%
  \BibitemOpen
  \bibfield  {author} {\bibinfo {author} {\bibfnamefont {E.~L.}\ \bibnamefont
  {Saldin}}, \bibinfo {author} {\bibfnamefont {E.~A.}\ \bibnamefont
  {Schneidmiller}}, \ and\ \bibinfo {author} {\bibfnamefont {M.}~\bibnamefont
  {Yurkov}},\ }\href@noop {} {\bibfield  {journal} {\bibinfo  {journal}
  {Nuclear Instruments and Methods in Physics Research Section A: Accelerators,
  Spectrometers, Detectors and Associated Equipment}\ }\textbf {\bibinfo
  {volume} {398}},\ \bibinfo {pages} {373} (\bibinfo {year}
  {1997})}\BibitemShut {NoStop}%
\bibitem [{\citenamefont {Guetg}\ \emph {et~al.}(2015)\citenamefont {Guetg},
  \citenamefont {Beutner}, \citenamefont {Prat},\ and\ \citenamefont
  {Reiche}}]{25}%
  \BibitemOpen
  \bibfield  {author} {\bibinfo {author} {\bibfnamefont {M.~W.}\ \bibnamefont
  {Guetg}}, \bibinfo {author} {\bibfnamefont {B.}~\bibnamefont {Beutner}},
  \bibinfo {author} {\bibfnamefont {E.}~\bibnamefont {Prat}}, \ and\ \bibinfo
  {author} {\bibfnamefont {S.}~\bibnamefont {Reiche}},\ }\href@noop {}
  {\bibfield  {journal} {\bibinfo  {journal} {Physical Review Special Topics -
  Accelerators and Beams}\ }\textbf {\bibinfo {volume} {18}} (\bibinfo {year}
  {2015})}\BibitemShut {NoStop}%
\bibitem [{\citenamefont {Guetg}\ \emph {et~al.}(2018)\citenamefont {Guetg},
  \citenamefont {Lutman}, \citenamefont {Ding}, \citenamefont {Maxwell},
  \citenamefont {Decker}, \citenamefont {Bergmann},\ and\ \citenamefont
  {Huang}}]{35}%
  \BibitemOpen
  \bibfield  {author} {\bibinfo {author} {\bibfnamefont {M.~W.}\ \bibnamefont
  {Guetg}}, \bibinfo {author} {\bibfnamefont {A.~A.}\ \bibnamefont {Lutman}},
  \bibinfo {author} {\bibfnamefont {Y.}~\bibnamefont {Ding}}, \bibinfo {author}
  {\bibfnamefont {T.~J.}\ \bibnamefont {Maxwell}}, \bibinfo {author}
  {\bibfnamefont {F.~J.}\ \bibnamefont {Decker}}, \bibinfo {author}
  {\bibfnamefont {U.}~\bibnamefont {Bergmann}}, \ and\ \bibinfo {author}
  {\bibfnamefont {Z.}~\bibnamefont {Huang}},\ }\href@noop {} {\bibfield
  {journal} {\bibinfo  {journal} {Physical Review Letters}\ }\textbf {\bibinfo
  {volume} {120}},\ \bibinfo {pages} {014801} (\bibinfo {year}
  {2018})}\BibitemShut {NoStop}%
\bibitem [{\citenamefont {Fl{\"o}ttmann}\ \emph {et~al.}(2011)\citenamefont
  {Fl{\"o}ttmann} \emph {et~al.}}]{53}%
  \BibitemOpen
  \bibfield  {author} {\bibinfo {author} {\bibfnamefont {K.}~\bibnamefont
  {Fl{\"o}ttmann}} \emph {et~al.},\ }\href@noop {} {\enquote {\bibinfo {title}
  {Astra: A space charge tracking algorithm},}\ } (\bibinfo {year}
  {2011})\BibitemShut {NoStop}%
\bibitem [{\citenamefont {Borland}(2000)}]{54}%
  \BibitemOpen
  \bibfield  {author} {\bibinfo {author} {\bibfnamefont {M.}~\bibnamefont
  {Borland}},\ }\href@noop {} {\bibfield  {journal} {\bibinfo  {journal} {ANL
  Advanced Photon Source Report No. LS-287}\ }\textbf {\bibinfo {volume} {28}}
  (\bibinfo {year} {2000})}\BibitemShut {NoStop}%
\bibitem [{\citenamefont {Reiche}(1999)}]{55}%
  \BibitemOpen
  \bibfield  {author} {\bibinfo {author} {\bibfnamefont {S.}~\bibnamefont
  {Reiche}},\ }\href@noop {} {\bibfield  {journal} {\bibinfo  {journal}
  {Nuclear Instruments and Methods in Physics Research Section A: Accelerators,
  Spectrometers, Detectors and Associated Equipment}\ }\textbf {\bibinfo
  {volume} {429}},\ \bibinfo {pages} {243} (\bibinfo {year}
  {1999})}\BibitemShut {NoStop}%
\bibitem [{\citenamefont {Das}\ and\ \citenamefont {Dennis}(1998)}]{41}%
  \BibitemOpen
  \bibfield  {author} {\bibinfo {author} {\bibfnamefont {I.}~\bibnamefont
  {Das}}\ and\ \bibinfo {author} {\bibfnamefont {J.~E.}\ \bibnamefont
  {Dennis}},\ }\href@noop {} {\bibfield  {journal} {\bibinfo  {journal} {SIAM
  Journal on Optimization}\ }\textbf {\bibinfo {volume} {8}},\ \bibinfo {pages}
  {631} (\bibinfo {year} {1998})}\BibitemShut {NoStop}%
\bibitem [{\citenamefont {Zhao}\ \emph {et~al.}(2017)\citenamefont {Zhao},
  \citenamefont {Wang}, \citenamefont {Gu}, \citenamefont {Yin}, \citenamefont
  {Gu}, \citenamefont {Leng},\ and\ \citenamefont {Liu}}]{43}%
  \BibitemOpen
  \bibfield  {author} {\bibinfo {author} {\bibfnamefont {Z.}~\bibnamefont
  {Zhao}}, \bibinfo {author} {\bibfnamefont {D.}~\bibnamefont {Wang}}, \bibinfo
  {author} {\bibfnamefont {Q.}~\bibnamefont {Gu}}, \bibinfo {author}
  {\bibfnamefont {L.}~\bibnamefont {Yin}}, \bibinfo {author} {\bibfnamefont
  {M.}~\bibnamefont {Gu}}, \bibinfo {author} {\bibfnamefont {Y.}~\bibnamefont
  {Leng}}, \ and\ \bibinfo {author} {\bibfnamefont {B.}~\bibnamefont {Liu}},\
  }\href@noop {} {\bibfield  {journal} {\bibinfo  {journal} {Applied Sciences}\
  }\textbf {\bibinfo {volume} {7}},\ \bibinfo {pages} {607} (\bibinfo {year}
  {2017})}\BibitemShut {NoStop}%
\bibitem [{\citenamefont {Huang}\ \emph {et~al.}(2014)\citenamefont {Huang},
  \citenamefont {Ding}, \citenamefont {Huang},\ and\ \citenamefont
  {Qiang}}]{36}%
  \BibitemOpen
  \bibfield  {author} {\bibinfo {author} {\bibfnamefont {S.}~\bibnamefont
  {Huang}}, \bibinfo {author} {\bibfnamefont {Y.}~\bibnamefont {Ding}},
  \bibinfo {author} {\bibfnamefont {Z.}~\bibnamefont {Huang}}, \ and\ \bibinfo
  {author} {\bibfnamefont {J.}~\bibnamefont {Qiang}},\ }\href@noop {}
  {\bibfield  {journal} {\bibinfo  {journal} {Phys. Rev. ST Accel. Beams}\
  }\textbf {\bibinfo {volume} {17}} (\bibinfo {year} {2014})}\BibitemShut
  {NoStop}%
\bibitem [{\citenamefont {Huang}\ \emph {et~al.}(2017)\citenamefont {Huang},
  \citenamefont {Ding}, \citenamefont {Feng}, \citenamefont {Hemsing},
  \citenamefont {Huang}, \citenamefont {Krzywinski}, \citenamefont {Lutman},
  \citenamefont {Marinelli}, \citenamefont {Maxwell},\ and\ \citenamefont
  {Zhu}}]{37}%
  \BibitemOpen
  \bibfield  {author} {\bibinfo {author} {\bibfnamefont {S.}~\bibnamefont
  {Huang}}, \bibinfo {author} {\bibfnamefont {Y.}~\bibnamefont {Ding}},
  \bibinfo {author} {\bibfnamefont {Y.}~\bibnamefont {Feng}}, \bibinfo {author}
  {\bibfnamefont {E.}~\bibnamefont {Hemsing}}, \bibinfo {author} {\bibfnamefont
  {Z.}~\bibnamefont {Huang}}, \bibinfo {author} {\bibfnamefont
  {J.}~\bibnamefont {Krzywinski}}, \bibinfo {author} {\bibfnamefont {A.~A.}\
  \bibnamefont {Lutman}}, \bibinfo {author} {\bibfnamefont {A.}~\bibnamefont
  {Marinelli}}, \bibinfo {author} {\bibfnamefont {T.~J.}\ \bibnamefont
  {Maxwell}}, \ and\ \bibinfo {author} {\bibfnamefont {D.}~\bibnamefont
  {Zhu}},\ }\href@noop {} {\bibfield  {journal} {\bibinfo  {journal} {Physical
  Review Letters}\ }\textbf {\bibinfo {volume} {119}},\ \bibinfo {pages}
  {154801} (\bibinfo {year} {2017})}\BibitemShut {NoStop}%
\bibitem [{\citenamefont {Piot}\ \emph {et~al.}(2012)\citenamefont {Piot},
  \citenamefont {Behrens}, \citenamefont {Gerth}, \citenamefont {Dohlus},
  \citenamefont {Lemery}, \citenamefont {Mihalcea}, \citenamefont {Stoltz},\
  and\ \citenamefont {Vogt}}]{38}%
  \BibitemOpen
  \bibfield  {author} {\bibinfo {author} {\bibfnamefont {P.}~\bibnamefont
  {Piot}}, \bibinfo {author} {\bibfnamefont {C.}~\bibnamefont {Behrens}},
  \bibinfo {author} {\bibfnamefont {C.}~\bibnamefont {Gerth}}, \bibinfo
  {author} {\bibfnamefont {M.}~\bibnamefont {Dohlus}}, \bibinfo {author}
  {\bibfnamefont {F.}~\bibnamefont {Lemery}}, \bibinfo {author} {\bibfnamefont
  {D.}~\bibnamefont {Mihalcea}}, \bibinfo {author} {\bibfnamefont
  {P.}~\bibnamefont {Stoltz}}, \ and\ \bibinfo {author} {\bibfnamefont
  {M.}~\bibnamefont {Vogt}},\ }\href@noop {} {\bibfield  {journal} {\bibinfo
  {journal} {Physical Review Letters}\ }\textbf {\bibinfo {volume} {108}},\
  \bibinfo {pages} {034801} (\bibinfo {year} {2012})}\BibitemShut {NoStop}%
\bibitem [{\citenamefont {Gao}\ \emph {et~al.}(2018)\citenamefont {Gao},
  \citenamefont {Ha}, \citenamefont {Jing}, \citenamefont {Antipov},
  \citenamefont {Power}, \citenamefont {Conde}, \citenamefont {Gai},
  \citenamefont {Chen}, \citenamefont {Shi},\ and\ \citenamefont
  {Wisniewski}}]{39}%
  \BibitemOpen
  \bibfield  {author} {\bibinfo {author} {\bibfnamefont {Q.}~\bibnamefont
  {Gao}}, \bibinfo {author} {\bibfnamefont {G.}~\bibnamefont {Ha}}, \bibinfo
  {author} {\bibfnamefont {C.}~\bibnamefont {Jing}}, \bibinfo {author}
  {\bibfnamefont {S.~P.}\ \bibnamefont {Antipov}}, \bibinfo {author}
  {\bibfnamefont {J.~G.}\ \bibnamefont {Power}}, \bibinfo {author}
  {\bibfnamefont {M.}~\bibnamefont {Conde}}, \bibinfo {author} {\bibfnamefont
  {W.}~\bibnamefont {Gai}}, \bibinfo {author} {\bibfnamefont {H.}~\bibnamefont
  {Chen}}, \bibinfo {author} {\bibfnamefont {J.}~\bibnamefont {Shi}}, \ and\
  \bibinfo {author} {\bibfnamefont {E.~E.}\ \bibnamefont {Wisniewski}},\
  }\href@noop {} {\bibfield  {journal} {\bibinfo  {journal} {Physical Review
  Letters}\ }\textbf {\bibinfo {volume} {120}},\ \bibinfo {pages} {114801}
  (\bibinfo {year} {2018})}\BibitemShut {NoStop}%
\end{thebibliography}%
\end{document}